\documentclass{elsarticle} 
\usepackage[margin=2.5cm]{geometry}
\usepackage{graphicx}\graphicspath{{Figs/}}
\usepackage[center]{subfigure}
\usepackage{epstopdf} 
\usepackage{amsmath,amssymb,amsfonts}
\usepackage{bm}
\begin{document}
\begin{frontmatter}
\title{A meshfree formulation for large deformation analysis of flexoelectric structures accounting for the surface effects}
\author[X,N1]{Xiaoying Zhuang\corref{cor2}}
\ead{zhuang@ikm.uni-hannover.de}
\cortext[cor2]{Corresponding author}
\author[N1]{S.S.Nanthakumar}
\author[WE]{Timon Rabczuk\corref{cor1}}
\ead{timon.rabczuk@uni-weimar.de}
\cortext[cor1]{Corresponding author}

\address[X] {College of Civil Engineering, Tongji University, 1239 Siping Road, 200092 Shanghai, China.}
\address[N1]{Institute of Continnum Mechanics, Leibniz University Hannover, Appelstrasse 11A, D-30167 Hannover Germany.}
\address[WE]{Institute of Structural Mechanics, Bauhaus University Weimar,  Marienstrasse 5, 99423 Weimar, Germany.}
\begin{abstract}
In this work, we present a compactly supported radial basis function (CSRBF) based meshfree method to analyse geometrically nonlinear flexoelectric nanostructures considering surface effects. Flexoelectricity is the polarization of dielectric materials due to the gradient of strain, which is different from piezoelectricity in which polarization is dependent linearly on strain. The surface effects gain prominence as the size of the structure tends to nanoscale and so their consideration is inevitable when flexoelectric nanostructures are analysed. First, the proposed meshfree formulation is validated and the influence of nonlinear strain terms on the energy conversion ability of flexoelectric beams made of a non-piezoelectric material like cubic Strontium Titanate is studied. Subsequently, the meshfree formulation for nonlinear flexoelectricity is extended to include nonlinear surface effects. It is determined that the surface effects can have notable influence on the output flexoelec
 tric voltage of nano-sized cantilever structures in the nonlinear regime.
\end{abstract}
\end{frontmatter}
\textbf{Keywords:}  Meshfree method; Nolinear Flexoelectricity; Geometric nonlinearity; Surface effects

\section{Introduction}
Flexoelectricity is the generation of electric polarization under mechanical strain gradient or mechanical deformation due to electric field gradient (converse flexo). It is a more general phenomenon than the linear change in polarization due to stress, known as the piezoelectric effect. Flexoelectric polarization is restricted not only to non-centrosymmetric crystals eventually opening up possibilities for nontoxic electromechanical materials for biomedical application. 

Piezoelectricity can be characterised by a third rank tensor and is observed only in non-centrosymmetric crystals (21 types). In contrast, flexoelectricity can be mathematically defined by a fourth order tensor and can be observed in materials of any symmetry (32). The reason behind is that the homogeneous strain relies on lack of symmetry of materials for polarization, on the other hand, the strain gradient can break the local centrosymmetry of materials inducing polarization. The strain gradient scales with the size of specimen leading to the possibility of significant flexoelectric effect at the length scale of nanometers. Piezoelectricity exists only below Curie temperature, while flexoelectricity being symmetry independent does not have a temperature constraint \cite{mbarki}. The high energy conversion ability of piezoelectric materials makes them the prominent constituent in several micro~\cite{priya} and nano-sized~\cite{wang_zl} energy harvesters developed. While, the recent 
 researches show the possibility of an energy harvester made of non-piezoelectric materials exploiting flexoelectricity \cite{sharma}. Nanoelectromechanical systems like actuators \cite{bhaskar} are fabricated using non-piezoelectric materials like Strontium Titanate and are shown to produce curvature/electric field ratio of 3.33 $MV^{-1}$ comparable to the ratio of 5.2 $MV^{-1}$ in piezoelectric Lead Zirconium Titanate bimorph. The flexoelectricity also offers the advantage of choosing Lead-free materials as constituents in sensors, actuators and energy harvesters.

The theory of flexoelectricity was first identified way back in the 1960s by Mashkevich and Tolpygo \cite{mashkevich}, followed later by the work of Tagantsev~\cite{Tagantsev} in which bulk and surface mechanisms that can cause polarization due to strain gradient were determined. Meanwhile, Kogan~\cite{Kogan} made a theoretical estimate of the flexoelectric coefficient to be of the order of $e/a$ $\approx$ $10^{-9} C/m$, where $e$ is the electronic charge and $a$ is the lattice parameter. The series of experimental works by Cross and co-workers~\cite{ma_mnio,ma_bst,ma_bto} sparked interest over the potential of flexoelectric materials as a substitute to piezoelectric materials. These experimental studies on ceramics with cubic symmetry like Barium Strontium Titanate (BST) and Barium Titanate (BTO) revealed higher values of peak flexoelectric coefficients in the range of 50 $\mu C/m$. Atomistically, Maranganti \textit{et al}.~\cite{margaranti:2006} determined the flexoelectric coeffic
 ients of several ferroelectric and non-ferroelectric crystals. Though there is discrepancy between theoretical and experimental flexoelectric coefficients of Barium Titanate (BTO), the theoretical estimations are of the same order compared to the experimental results of Zubko \textit{et al}.~\cite{Zubko} for Strontium Titanate (STO) crystals.  Several works are available in literature that presents analytically derived electro-elastic field for nanobeams and nanowires having flexoelectric effect and surface effects~\cite{jiang_beam,Liu,zhang}. 
 electric coefficients are set to be the minimum.
Nevertheless, the analytical solutions are applicable only to simplified one-dimensional models and so numerical  methods to analyse flexoelectric structures are required. 

Conventional finite element method (FEM) cannot be utilised for analysing flexoelectric structures as the fourth order partial differential equations governing flexoelectricity necessitates $C^1$ continuity of displacement field. Phase field modelling of flexoelectricity in an epitaxial thin film made of Barium Titanate is presented by Chen \textit{et al}. \cite{Chen}, followed by which analysis of a two phase system is performed \cite{Chen1}. Meshfree shape functions offer the advantage of having higher order continuity, making them a favourable class of numerical methods to analyse flexoelectric structures. A numerical approach to analyse two and three dimensional truncated pyramid shaped  structure due to flexoelectricity utilising local maximum entropy (LME) meshfree method is presented by Abdollahi \textit{et al}.~\cite{amir,amir3d}. Ghasemi \textit{et al}.~\cite{hamid} proposed an IGA formulation exploiting the higher order continuity of NURBS shape functions. In \cite{nantha:f
 lexo,Mao_FE,Shen_FE}, mixed FE formulations are proposed for analysis of two dimensional flexoelectric structures. Though the mixed FE formulation requires only $C^0$ continuity, the number of nodal DOFs required is much higher. For example, in the flexoelectric element proposed in~\cite{nantha:flexo}, degrees of freedom in the corner nodes are two displacement DOFs, four displacement gradient DOFs, one electric potential DOF and four  Lagrange multiplier DOFs. It is to be noted that in the work of Nanthakumar \textit{et al.}~\cite{nantha:flexo} flexoelectric nanobeams with surface effects, made of Barium Titanate, are analysed and optimized using the mixed FE formulation. Also there are computational works available in literature that specifically analyse nanobeams with surface effects~\cite{nantha:CM,nantha:JMPS,farsad}, in which extended finite element method is the numerical method adopted. The nonlinear electro-elasticity of soft dielectrics combined with flexoelectricity is an
 alysed by Yvonnet \textit{et al}.~\cite{yvonnet:nl}, adopting finite element discretization ($C^1$ Argyris triangular elements) and consistent linearizations. As a shortcoming the authors have stated that due to instability, the utilised formulation could not simulate the entire nonlinear range. 

Motivated by all these works on flexoelectricity, in the present work, a compactly supported radial basis function (CSRBF) based meshfree formulation is proposed to analyse flexoelectric beams subjected to large deformation considering surface effects. Though there are works available in literature on analysing flexoelectric structures using a meshfree method \cite{amir,amir3d}, to the best of our knowledge this is the first work on a meshfree formulation to handle nonlinearity in flexoelectric nanostructures accounting for surface effects. The meshfree shape functions with higher order continuity are advantageous compared to complex mixed FE \cite{nantha:flexo} formulations, mainly because the meshfree formulation requires the discretization of 'only' displacement and electric potential fields.

The outline of the paper is as follows: Section~\ref{sec2} presents the governing equations of flexoelectricity. Section~\ref{sec3} describes the meshfree formulation for flexoelectricity including surface elasticity and surface piezoelectricity. Linearization of the weak form and subsequent meshfree discretization is shown in Section~\ref{sec:4}. Finally, numerical examples on analysis of two-dimensional flexoelectric structures with surface effects considering geometric nonlinearity are presented in Section~\ref{sec5}.

\section{Governing equations of flexoelectricity  with surface effects}
\label{sec2}
The mathematical modelling of flexoelectricity is based on the extended linear theory of piezoelectricity with additional strain gradient terms. A general internal energy density function, U involving strain energy, electrostatic energy and terms including strain gradient is presented by Shen~\textit{et al.}~\cite{shen:2010}. The internal energy density function, $U$ is as follows,
\begin{equation}
\begin{split}
& U = U_{b} + U_{s}\\
& U_{b} = \frac{1}{2} \bm{\varepsilon}:\mathbf{C}:\bm{\varepsilon} - \mathbf{E} \cdot \mathbf{e} : \bm{\varepsilon} - \mathbf{E} \cdot \bm{\mu} \vdots \bm{\eta} - \frac{1}{2} \bm{E} \cdot \bm{\kappa} \cdot \bm{E}+\frac{1}{2}\bm{\eta}\vdots\bm{g}\vdots\bm{\eta} \\
& U_{s} = U_{s0} + \bm{\tau^s} : \bm{\varepsilon^s} + \bm{\omega^s} \cdot \mathbf{E^s} + \frac{1}{2} \bm{\varepsilon^s}:\mathbf{C^s}:\bm{\varepsilon^s} -
\mathbf{E^s} \cdot \mathbf{e^s} : \bm{\varepsilon^s} - \frac{1}{2} \bm{E^s} \cdot \bm{\kappa^s} \cdot \bm{E^s} 
\label{enthalpy}
\end{split}
\end{equation}
where ${U_b}$ and ${U_s}$ are the bulk and surface energy density functions respectively. $U_{s0}$ is the surface free energy density. $\bm{\varepsilon}$ is the linear strain tensor, $\bm{E}$ is the electric field tensor, $\bm{\varepsilon^s}$ and $\bm{E^s}$ are their corresponding surface counterparts. $\bm{\tau_s}$ and $\bm{\omega_s}$ are the residual surface stress and residual surface electric displacements respectively. $\bm{\eta}$ is the strain gradient tensor. $\bm{C}$ and $\bm{C^s}$ are the fourth order bulk and surface stiffness tensors, $\bm{e}$ and $\bm{e^s}$ are the third order bulk and surface piezoelectric coupling tensors, $\bm{\kappa}$ and $\bm{\kappa^s}$ are the bulk and surface dielectric permittivity tensors respectively. $\bm{g}$ is the sixth order strain gradient elasticity tensor. $\bm{\mu}$ is the fourth order flexoelectric tensor which represents combination of (a) strain-polarization gradient coupling and (b) strain gradient-polarization coupling. 

The physical stress, $\bm{\sigma}$ and electric displacement, $\bm{D}$ can be obtained from the bulk energy density function as,
\begin{equation}
\sigma_{ij}= \frac{\partial U_b}{\partial \varepsilon_{ij}} - \left(\frac{\partial U_b}{\partial \eta_{ijk}}\right)_{,k} = C_{ijkl} \varepsilon_{kl} - e_{ijk} E_{k} + \mu_{ijkl}E_{k,l}-g_{ijklmn}\eta_{lmn,k}
\end{equation}
\begin{equation}
D_i = -\frac{\partial U_b}{\partial E_{i}} = e_{ijk}\varepsilon_{jk} + \mu_{ijkl}\varepsilon_{jk,l} + \kappa_{ij}E_j 
\end{equation}\\
The surface mechanical stress, $\bm{\sigma^s}$ and surface electric displacement, $\bm{D^s}$ can be obtained from the surface energy density function as,\\
\begin{equation}
\sigma^s_{ij}= \frac{\partial U_s}{\partial \varepsilon^s_{ij}} = \tau^s_{ij} + C^s_{ijkl} \varepsilon^s_{kl} - e^s_{ijk} E^s_{k}
\end{equation}
\begin{equation}
D^s_i = -\frac{\partial U_s}{\partial E^s_{i}} = -\omega^s_i + e^s_{ijk}\varepsilon^s_{jk} + \kappa^s_{ij}E^s_j 
\end{equation}\\ 
The total potential energy, $\Pi$ can be written in terms of internal energy in the bulk, $\Pi_{bulk}$, internal energy in the surface, $\Pi_s$ and work done by external forces, $\Pi_{ext}$ as,
\begin{equation}
\Pi =  \Pi_{bulk} + \Pi_{s} - \Pi_{ext}
\label{PE}
\end{equation}
where,
\begin{align}
\Pi_{bulk}&=\int_\Omega U_b\;d\Omega\\
\Pi_{s}&=\int_\Gamma U_s\;d\Gamma\\
\Pi_{ext}&=\int_{\Gamma_u}\bm{u}\cdot\bm{t}\;d\Gamma_u+\int_\Omega\bm{u}\cdot\bm{b}\;d\Omega-
\int_{\Gamma_\phi}\phi q\;d\Gamma_\phi
\end{align}
Here, $\bm{u}$ and $\phi$ denote mechanical displacement and electric potential respectively. $\bm{t}$ is the surface traction on $\Gamma_u$, $\bm{b}$ is the prescribed body force and $q$ is the surface charge density on $\Gamma_\phi$. $\Gamma_u$ and $\Gamma_\phi$ are the Neumann boundary for mechanical displacement and electric potential respectively.\\
The weak form of the equilibrium equations can be obtained by finding $\mathbf{u}\in\{\mathbf{u}=\bar{\mathbf{u}}\quad\text{on }\mathbf{\Gamma}^d_u, \mathbf{u}\in H^2(\Omega)\}$ and $\bm{\phi}\in\{\bm{\phi}=\bar{\bm{\phi}}\quad\text{on }\mathbf{\Gamma}^d_{\phi}, \bm{\phi}\in H^2(\Omega)\}$ such that 
\begin{equation}
\begin{split}
&\delta \Pi = 0 \implies \\
&\int_{\Omega}\bm{\varepsilon}(\delta\mathbf{u}):\mathbf{C}:\bm{\varepsilon}(\mathbf{u})\,d\Omega-\int_{\Omega} \bm{\varepsilon}(\delta\mathbf{u}) : \mathbf{e} \cdot \bm{E}(\bm{\phi})  \,d\Omega\\
&-\int_{\Omega} \bm{E}(\delta\bm{\phi})\cdot\mathbf{e}: \bm{\varepsilon}(\mathbf{u})\,d\Omega-\int_{\Omega} \bm{\eta}(\delta\mathbf{u})\vdots\bm{\mu}\cdot \bm{E}(\bm{\phi})\,d\Omega \\
&-\int_{\Omega} \bm{E}(\delta\bm{\phi})\cdot\bm{\mu}\vdots\bm{\eta}(\mathbf{u})\,d\Omega-\int_{\Omega} \bm{E}(\delta\bm{\phi}) \cdot \bm{\kappa} \cdot \bm{E}(\bm{\phi}) \,d\Omega\\
&+\int_\Omega \bm{\eta(\delta u)}\vdots\bm{g}\vdots \bm{\eta(u)}\;d\Omega + \int_{\Gamma}\bm{\varepsilon}^s(\delta\mathbf{u}):\bm{\tau^s}\,d\Gamma \\ 
&+ \int_{\Gamma}\bm{E}^s(\delta\bm{\phi}) \cdot \bm{\omega^s} \,d\Gamma + \int_{\Gamma}\bm{\varepsilon^s}(\delta\mathbf{u}):\mathbf{C^s}:\bm{\varepsilon^s}(\mathbf{u})\,d\Gamma \\
&-\int_{\Gamma} \bm{\varepsilon^s}(\delta\mathbf{u}) : \mathbf{e^s} \cdot \bm{E^s}(\bm{\phi})  \,d\Gamma 
-\int_{\Gamma} \bm{E^s}(\delta\bm{\phi})\cdot\mathbf{e^s}: \bm{\varepsilon^s}(\mathbf{u})\,d\Gamma \\
&-\int_{\Gamma} \bm{E^s}(\delta\bm{\phi}) \cdot \bm{\kappa^s} \cdot \bm{E^s}(\bm{\phi}) \,d\Gamma
=\int_{\Gamma_{u}}\delta\mathbf{u}\cdot\bar{\mathbf{t}}\,d\Gamma_u
 +\int_{\Omega}\delta\mathbf{u}\cdot\mathbf{b}\,d\Omega
-\int_{\Gamma_\phi}\delta\bm{\phi} \, {q}\,d\Gamma_\phi
\end{split}
\label{weak form}
\end{equation}
for all $\delta\mathbf{u}\in\{\delta\mathbf{u}=0\quad\text{on }{\Gamma}^d_u, \delta\mathbf{u}\in H^2(\Omega)\}$ and $\delta\bm{\phi}\in\{\delta\bm{\phi}=0\quad\text{on }{\Gamma}^d_\phi, \delta\bm{\phi}\in H^2(\Omega)\}$. $\Gamma^d_u$ and $\Gamma^d_\phi$ are the Dirichlet boundary for mechanical displacement and electric potential respectively.

\section{Mesh free formulation for flexoelectricity}
\label{sec3}
The numerical discretization of the governing partial differential equations of flexoelectricity requires  $C^1$ continuous basis functions for a Galerkin method. In the present work, we utilize a meshfree method with compactly supported radial basis function (CSRBF)  shape functions. Popular radial basis functions \cite{liu2009} include the Multi-Quadrics, Gaussian and Thin Plate Splines. These radial basis functions are globally supported and their accuracy highly depends on the condition number of the collocation matrix. However, the collocation matrix will be a sparse matrix, well conditioned and compactly supported if we adopt CSRBF shape functions. The Wendland type CSRBFs with $C^2$ and $C^4$ continuity proposed in \cite{wendland:1995} are,
\begin{equation}
f(r(x,y))=max\left\lbrace 0, (1-r)^4 \right\rbrace (4r+1)\;\; \in \;\;C^2
\label{csrbf1}
\end{equation}
\begin{equation}
f(r(x,y))=max\left\lbrace 0, (1-r)^6 \right\rbrace (35r^2+18r+3)\;\; \in \;\;C^4
\label{csrbf2}
\end{equation}
where $r(x,y)$ is given by,
\begin{equation}
r_i(x,y)= \frac{d_i}{R} = \frac{\sqrt{(x-x_i)^2 + (y-y_i)^2}}{R}
\end{equation}
where $d_i$ is the distance of a point of interest $(x,y)$ from a knot at $(x_i,y_i)$. The dimension of support domain, $R$ is given by $ R = \alpha d_c $, where $\alpha$ is the shape parameter and $d_{c}$ is the average nodal spacing. It is to be noted that the value of $r_i$ lies between 0 and 1.\\
An approximation for a general function can be written as
\begin{equation}
{u^h(\texttt{x})} = \bm{f^T(\texttt{x})}\,\bm{a} + \bm{p^T(\texttt{x})}\,\bm{b}
\label{eq:approx}
\end{equation}
where $\bm{f(\texttt{x})}$ and $\bm{a}$ denote the vector of CSRBF and expansion coefficients respectively,
\begin{equation}
\bm{f^T(\texttt{x})}=[f_1(\texttt{x}),f_2(\texttt{x}),...f_n(\texttt{x})]
\label{Eq:f}
\end{equation}
\begin{equation}
\bm{a^T}=[a_1,a_2,...a_n].
\label{Eq:a}
\end{equation}
In  Equations~\ref{Eq:f} and~\ref{Eq:a}, the variable $n$ stands for the number of nodes in
the support domain of the point of interest. Here, $\bm{p(\texttt{x})}$ and $\bm{b}$ are the vector of polynomial basis functions and coefficients respectively,
\begin{equation}
\bm{p^T(\texttt{x})}=[p_1(\texttt{x}),p_2(\texttt{x}),...p_m(\texttt{x})]
\label{eq:poly}
\end{equation}
\begin{equation}
\bm{b^T}=[b_1,b_2,...b_m].
\label{Eq:b}
\end{equation}
In Equations~\eqref{eq:poly} and~\eqref{Eq:b}, the variable $m$ stands for the number of terms of polynomial basis.
The coefficient vectors $\bm{a}$ and $\bm{b}$ can be obtained by solving the following algebraic equations
\begin{equation}
 \left[\begin{array}{c c}
\bm{A} & \bm{P_m}\\ \bm{P_m^T} & 0\\
\end{array}\right] \left\lbrace\begin{array}{c}\bm{a}
\\
\bm{b}\end{array}\right\rbrace = \left\lbrace\begin{array}{c}\bm{U}
\\
0\end{array}\right\rbrace \ , 
\end{equation}
where $\bm{U}$ is a vector of nodal values of function $u^h(\texttt{x})$
and matrices $\bm{A}$ and $\bm{P_m}$ are,
\begin{equation}
\begin{split}
&\mathbf{A}=\left[\begin{array}{c c c}
f_1(\bm{\texttt{x}_1}) & \cdots & f_n(\bm{\texttt{x}_1})\\ \vdots & \ddots & \vdots\\f_1(\bm{\texttt{x}_n}) & \cdots & f_n(\bm{\texttt{x}_n})
\end{array}\right]
\end{split}
\end{equation}
\begin{equation}
\begin{split}
&\mathbf{P_m}=\left[\begin{array}{c c c}
p_1(\bm{\texttt{x}_1}) & \cdots & p_m(\bm{\texttt{x}_1})\\ \vdots & \ddots & \vdots\\p_1(\bm{\texttt{x}_n}) & \cdots & p_m(\bm{\texttt{x}_n})
\end{array}\right]
\end{split} \ . 
\end{equation}
From Equation~\eqref{eq:approx}, interpolation of the nodal function
values, $\bm{U}$, at any point of interest, $x$ can be written as,
\begin{equation}
\begin{split}
u^h(\texttt{x}) & = [\bm{f^T(\texttt{x})\;S_a}+\bm{p^T(\texttt{x})\;S_b}]\;\bm{U}\\
& = \bm{N(\texttt{x})} \bm{U}
\end{split}
\end{equation}
As a result, the meshfree CSRBF based shape function, $\bm{N(\texttt{x})}$ is given by,
\begin{equation}
\bm{N(\texttt{x})}= \bm{f^T(\texttt{x})\;S_a}+\bm{p^T(\texttt{x})\;S_b}
\end{equation}\\
where $\bm{S_a}\;=\;\bm{A^{-1}}[1-\bm{P_m}\,\bm{S_b}]$ and $\bm{S_b}\;=\;[\bm{P_m^T\,A^{-1}\,P_m}]^{-1}\bm{P_m^T\,A^{-1}}$.\\
\\
The polynomial basis functions of linear order are added to the radial basis functions in order to ensure that the shape functions possess $C^1$ consistency. The vector $\bm{p(\texttt{x})}$ given in Equation~\eqref{eq:poly} can be rewritten such that $m=3$ as,
\begin{equation}
\bm{p^T(\texttt{x})}=[1\;x\;y]
\end{equation}
The discrete form of the weak formulation in Equation~\eqref{weak form} using the meshfree shape functions is as follows,
\begin{equation}
\begin{split}
& \bm{\delta u}^T \left( \int_\Omega \bm{B_u}^T \bm{C} \bm{B_u} d\Omega \right) \bm{u} +  \bm{\delta u}^T \left( \int_\Omega \bm{B_u}^T \bm{e}^T \bm{B_\phi} d\Omega \right) \bm{\phi}\; + \\ 
& \bm{\delta \phi}^T \left( \int_\Omega \bm{B_\phi}^T \bm{e} \bm{B_u} d\Omega \right) \bm{u}\;+ \bm{\delta u}^T \left( \int_\Omega \bm{H_u}^T \bm{\mu}^T \bm{B_\phi} d\Omega \right) \bm{u} + \\
& \bm{\delta \phi}^T \left( \int_\Omega \bm{B_\phi}^T  \bm{\mu} \bm{H_u} d\Omega \right) \bm{\phi}
- \bm{\delta\phi}^T \left( \int_\Omega  \bm{B_\phi}^T \bm{\kappa} \bm{B_\phi} d\Omega \right) \bm{\phi}\;+ \\ 
& \bm{\delta u}^T \left( \int_\Omega \bm{H_u}^T \bm{g} \bm{H_u} d\Omega \right) \bm{u} + \bm{\delta u}^T \left( \int_\Gamma \bm{B_u}^T \bm{M_p}^T\bm{\tau^s} d\Gamma \right)-\\
& \bm{\delta \phi}^T \left( \int_\Gamma \bm{P}^T \bm{\omega^s} d\Gamma \right) + \bm{\delta u}^T \left( \int_\Gamma \bm{B_u}^T \bm{M_p}^T \bm{C^s} \bm{M_p} \bm{B_u} d\Gamma \right) \bm{u}\;-\\
& \bm{\delta u}^T \left( \int_\Gamma \bm{B_u}^T \bm{M_p}^T {\bm{e^s}}^T \bm{P} \bm{B_\phi} d\Gamma \right) \bm{\phi}  + \bm{\delta \phi}^T \left( \int_\Gamma \bm{B_\phi}^T \bm{P}^T \bm{e^s} \bm{M_p} \bm{B_u} d\Gamma \right) \bm{u}-\;\\
& \bm{\delta \phi}^T \left( \int_\Gamma \bm{B_\phi}^T \bm{P}^T \bm{\kappa^s} \bm{P}\bm{B_\phi} d\Gamma \right) \bm{\phi}\\
&  = \bm{\delta u}\int_{\Gamma_u} \bm{N}^T \bm{\bar{t}} d\Gamma_u
+ \bm{\delta u}\int_\Omega \bm{N}^T \bm{b} d\Omega - \bm{\delta \phi}\int_{\Gamma_\phi} \bm{N}^T {q} d\Gamma_\phi
\end{split}
\label{Eq:alg_form}
\end{equation}
where $\bm{C}$, $\bm{e}$, $\bm{\mu}$, $\bm{\kappa}$ and $\bm{g}$ are the matrix form of the tensors ${C_{ijkl}}$, ${e_{ijk}}$, ${\mu_{ijkl}}$, ${\kappa_{ij}}$ and ${g_{ijklmn}}$ respectively and $\bm{C^s}$, $\bm{e^s}$, $\bm{\mu^s}$ and $\bm{\kappa^s}$ are the matrix form of the tensors ${C^s_{ijkl}}$, ${e^s_{ijk}}$, ${\mu^s_{ijkl}}$, and ${\kappa^s_{ij}}$ respectively. The gradient and Hessian matrices in Equation~\ref{Eq:alg_form} are defined as follows,
\begin{equation}
\bm{B_u} = \left[ \begin{array}{cc} N_{I,x} & 0 \\ 0 & N_{I,y} \\  N_{I,y} & N_{I,x} \end{array} \right] 
\end{equation}
\begin{equation}
\bm{B_\phi} = -\left[ \begin{array}{c} N_{I,x} \\ N_{I,y} \end{array} 
\right]
\end{equation}
\begin{equation}
\bm{H_u} = \left[ \begin{array}{cc} N_{I,xx} & 0 \\ 0 & N_{I,yx} \\ N_{I,yx} & N_{I,xx}\\
N_{I,xy} & 0\\
0 & N_{I,yy}\\
N_{I,yy} & N_{I,xy}
\end{array}
\right]
\end{equation}\\
where $I=1,2....n$, $n$  is the number of nodes in the support domain of the point of interest and this number can be different for different points of interest. The projection matrix is denoted as $\bm{M_P}$
\begin{equation}
\bm{M_P} = \left( \begin{array}{l}
 P^2_{11}\;\;\;\;\;\;\;\;\;P^2_{12}\;\;\;\;\;\;\;\;\;P_{11}P_{12} \\
P^2_{12}\;\;\;\;\;\;\;\;\;P^2_{22}\;\;\;\;\;\;\;\;\;P_{12}P_{22}\\
2P_{11}P_{12}\;\;2P_{12}P_{22}\;\;P^2_{12}+P_{11}P_{22} \end{array} \right)
\end{equation}
where the entries of  $\bm{M_P}$ are from  $\bm{P}$, the tangential projection tensor given by $\bm{I}-\mathbf{n}\otimes\mathbf{n}$. Here, $\bm{I}$ refers to identity matrix of rank 2 and $\bm{n}$ is the outward normal vector to the surface, $\Gamma$. The final system of equations can be written as follows,\\
\begin{equation}
\left[
\begin{array}{cc}
\bm{K_{uu}}+\bm{K_{uu}^s} & \bm{K_{u\phi}} +\bm{K_{u\phi}^s}\\
\bm{K_{\phi u}}+\bm{K_{\phi u}^s} & \bm{K_{\phi\phi}}+\bm{K_{\phi\phi}^s}
\end{array}
\right]
\left[
\begin{array}{c}
\bm{u} \\ \bm{\phi}
\end{array}
\right]
=
\left[
\begin{array}{c}
\bm{F_u}+\bm{F_u^s} \\ \bm{F_\phi}+\bm{F_\phi^s}
\end{array}
\right]
\label{algeqn}
\end{equation}
where 
\begin{equation} \label{eq:stiffnesses}
\begin{split}
\bm{K_{uu}} &=  \int_\Omega \bm{B_u}^T \mathbf{C} \bm{B_u} d\Omega +  \int_\Omega \bm{H_u}^T \bm{g} \bm{H_u} d\Omega \\ 
\bm{K_{u\phi}} &=  \int_\Omega \bm{B_u}^T \bm{e}^T \bm{B_\phi} d\Omega + \int_\Omega \bm{H_u}^T \bm{\mu}^T \bm{B_\phi} d\Omega  = \bm{K_{\phi u}}^T \\ 
\bm{K_{\phi \phi}} &=  -\int_\Omega \bm{B_\phi}^T \bm{\kappa} \bm{B_\phi} d\Omega \\
\bm{K_{uu}^s} &=  \int_\Gamma \bm{B_u}^T \bm{M_p}^T \bm{C^s} \bm{M_p} \bm{B_u} d\Gamma \\
\bm{K_{\phi u}^s} &=  \int_\Gamma \bm{B_\phi}^T \bm{P}^T \bm{e^s} \bm{M_p} \bm{B_u} d\Gamma  = {\bm{K_{u\phi} ^s}}^T \\ 
\bm{K_{\phi \phi}^s} &=  -\int_\Gamma \bm{B_\phi}^T \bm{P}^T\bm{\kappa^s}\bm{P} \bm{B_\phi} d\Gamma \\
\bm{F_u^s} &= \int_\Gamma \bm{B_u}^T \bm{M_p}^T\bm{\tau^s} d\Gamma \\
\bm{F_\phi^s} &= \int_\Gamma \bm{B_\phi}^T \bm{P}^T \bm{\omega^s} d\Gamma \\
\bm{F_u} & = \bm{\delta u}\int_{\Gamma_u} \bm{N}^T \bm{\bar{t}} d{\Gamma_u}
+ \bm{\delta u}\int_\Omega \bm{N}^T \bm{b} d\Omega \\
\bm{F_\phi} &= \bm{\delta \phi}\int_{\Gamma_{\phi}} \bm{N}^T \bm{q} d{\Gamma_\phi}\\
\end{split}
\end{equation}
\begin{equation}
\begin{split}
\bm{C}&=\left[ 
\begin{array}{ccc}
C_{11} & C_{12} & 0  \\
C_{12} & C_{22} & 0 \\
0 & 0 & C_{66}
\end{array}
\right] \\
\bm{\mu}&=\left[ 
\begin{array}{cccccc}
\mu_{11} & \mu_{12} & 0 & 0 & 0 & \mu_{44} \\
0 & 0 & \mu_{44} & \mu_{12} & \mu_{11} & 0 
\end{array}
\right] \\
\bm{e}&=\left[ 
\begin{array}{ccc}
0 & 0 & e_{15} \\
e_{31} & e_{33} & 0 
\end{array}
\right]\\
\bm{\kappa}&=\left[ 
\begin{array}{cc}
\kappa_{11}  & 0  \\
0 & \kappa_{22}
\end{array}
\right] \\
\end{split}
\end{equation}
\begin{equation}
\begin{split}
\bm{g}&={l_0}^2\left[ 
\begin{array}{cccccc}
C_{11} & C_{12} & 0 & 0 & 0 & 0 \\
C_{12} & C_{22} & 0 & 0 & 0 & 0 \\
0 & 0 & C_{66} & 0 & 0 & 0 \\
0 & 0 & 0 & C_{11} & C_{12} & 0\\
0 & 0 & 0 & C_{12} & C_{22} & 0 \\
0 & 0 & 0 & 0 & 0 & C_{66} \\
\end{array}
\right]
\end{split}
\label{Eq:g}
\end{equation}
In Equation~\ref{Eq:g}, the term $l_0$ is the length scale representing the size dependency of strain gradient effects.
\section{Mesh free formulation for flexoelectricity including geometric nonlinearity}
\label{sec:4}
In this section, the proposed meshfree formulation is extended to handle geometric nonlinearities in flexoelectric structures considering surface elasticity. Saint Venant-Kirchhoff material model is considered for flexoelectric solids, the internal energy density given in Equation~\ref{enthalpy} is modified as,\\
\begin{equation}
U = \frac{1}{2} \bm{S} : \bm{G} + \frac{1}{2} \bm{\tilde{S}} \vdots \bm{\tilde{G}} -\frac{1}{2} \bm{D} \cdot \bm{E} + \frac{1}{2} \bm{S^s} : \bm{G^s}
\end{equation}
The total potential energy, $\Pi$ is given by,
\begin{equation}
\begin{split}
\Pi &= \int_\Omega U d\Omega-\int_{\Gamma_u}\bm{u}\cdot\bm{t}\;d\Gamma_u-\int_\Omega\bm{u}\cdot\bm{b}\;d\Omega\\
&+\int_{\Gamma_\phi}\phi q\;d\Gamma_\phi
\end{split}
\label{Eq:Pi_nl}
\end{equation}
where $\bm{S}$ is the second Piola-Kirchhoff tensor, $\bm{\tilde{S}}$ is the double stress tensor and $\bm{D}$ is the electric displacement vector; $\bm{S^s}$ is the second Piola-Kirchhoff surface stress tensor; $\bm{G}$ is the Green Lagrange strain tensor and $\bm{\tilde{G}}$ is the gradient of the Green Lagrange strain tensor; $\bm{E}$ is the electric field vector and $\bm{G^s}$ is the Green Lagrange surface strain tensor. $\bm{u}$ and $\phi$ are mechanical displacement and electric potential respectively. $\bm{t}$ is the surface traction on $\Gamma_u$, $\bm{b}$ is the prescribed body force and $q$ is the surface charge density on $\Gamma_\phi$. $\Gamma_u$ and $\Gamma_\phi$ are the Neumann boundary for mechanical displacement and electric potential respectively.  

The constitutive equations of the assumed Saint-Venant Kirchhoff material model are as follows
\begin{equation}
\begin{split}
&\bm{S}=\bm{C}:\bm{G}-\bm{e}\cdot\bm{E}\\
&\bm{\tilde{S}}=-\bm{\mu}\cdot\bm{E}+\bm{g}\vdots\bm{\tilde{G}}\\
&\bm{D}=\bm{e}:\bm{G}+\bm{\mu}\vdots\bm{\tilde{G}}+\bm{\kappa}\cdot\bm{E}\\
\end{split}
\end{equation}
Taking the first variation of the total potential energy in Equation~\ref{Eq:Pi_nl} yields,
\begin{equation}
\begin{split}
\delta \Pi&=\int_{\Omega} \bm{S} : \bm{\delta G} d\Omega + \int_\Omega \bm{\tilde{S}} \vdots\bm{\delta \tilde{G}} d\Omega - \int_\Omega \bm{D} \cdot\bm{\delta E}d\Omega + \int_{\Gamma} \bm{S^s} : \bm{\delta G^s} d\Gamma \\
& - \int_{\Gamma_u} \bm{\delta u}\cdot\bm{t}   d\Gamma_u - \int_\Omega \bm{\delta u}\cdot\bm{b} d\Omega +\int_{\Gamma_\phi}\delta\bm{\phi} \, {q}\,d\Gamma_\phi = 0
\end{split}
\label{eq:work1}
\end{equation}
Each term in Equation~\ref{eq:work1} has to be linearized. The final expression obtained after linearizing each term in Equation~\ref{eq:work1} are subsequently presented. The intermediate steps are detailed in \ref{app1}. A total Lagrangian formulation is presented such that all the integrals are performed on the undeformed configuration and derivatives are with respect to the material coordinates.

The linearization of the term $\int_{\Omega} \bm{S} : \bm{\delta G} d\Omega$ in Equation~\ref{eq:work1}  can be obtained as
\begin{equation}
L\left[\int_\Omega \bm{S}:\bm{\delta G} d\Omega\right] = \int_\Omega \bm{\bar{S}}:\delta \bm{\bar{G}} d\Omega + \int_\Omega \Delta(\bm{S}:\bm{\delta G}) d\Omega
\label{eq:LT1}
\end{equation}
\begin{equation}
\begin{split}
\int_\Omega \Delta(\bm{S}:\bm{\delta G}) d\Omega &= \int_\Omega \bm{S}:\Delta(\bm{\delta G}) d\Omega + \int_\Omega \bm{\delta G}:\Delta\bm{S} d\Omega\\
&= \int_\Omega \bm{S}:\Delta\bm{(\delta G)} d\Omega + \int_\Omega \bm{\delta G}:\bm{C}:\Delta\bm{G} d\Omega - \int_\Omega \bm{\delta G}:\bm{e}\cdot\Delta\bm{E} d\Omega\\
&= \int_\Omega \bm{S}:[(\nabla_0 \bm{\delta u})^T (\nabla_0 (\Delta\bm{u}))] d\Omega + \int_\Omega \bm{\delta G}:\bm{C}:\Delta\bm{G} d\Omega- \int_\Omega \bm{\delta G}:\bm{e}\cdot\Delta\bm{E} d\Omega\\
\end{split} \ . 
\end{equation}

The linearization of the term $\int_\Omega \bm{\tilde{S}}\vdots\bm{\delta \tilde{G}} d\Omega$ in Equation~\ref{eq:work1} can be derived as follows,
\begin{equation}
L\left[\int_\Omega \bm{\tilde{S}}\vdots\bm{\delta \tilde{G}} d\Omega\right] = \int_\Omega \bm{\bar{\tilde{S}}}\vdots\bm{\delta \bar{\tilde{G}}} d\Omega + \int_\Omega \Delta(\bm{\tilde{S}}\vdots\bm{\delta \tilde{G}}) d\Omega
\label{eq:LT2}
\end{equation}
\begin{equation}
\begin{split}
\int_\Omega \Delta(\bm{\tilde{S}}\vdots\bm{\delta \tilde{G}}) d\Omega &= \int_\Omega \bm{\tilde{S}}\vdots \Delta(\bm{\delta \tilde{G}}) d\Omega + \int_\Omega \bm{\delta \tilde{G}}\vdots \Delta\bm{\tilde{S}} d\Omega\\
&= \int_\Omega \bm{\tilde{S}}\vdots \Delta(\bm{\delta \tilde{G}}) d\Omega - \int_\Omega \bm{\delta\tilde{G}} \vdots \bm{\mu}\cdot\Delta\bm{E} d\Omega \\
&= \int_\Omega \bm{\tilde{S}}\vdots[(\bm{\nabla_0^2 \delta u})(\bm{\nabla_0} \Delta \bm{u}) + (\bm{\nabla_0^2} \Delta \bm{u})(\bm{\nabla_0 \delta u})] d\Omega - \int_\Omega \bm{\delta \tilde{G}}\vdots \bm{\mu} \cdot\Delta\bm{E} d\Omega\\
\end{split}
\end{equation}

The linearization of the term $\int_\Omega \bm{D}\cdot\bm{\delta E} d\Omega$ in Equation~\ref{eq:work1} is as follows,\\
\begin{equation}
L\left[\int_\Omega \bm{D}\cdot\bm{\delta E} d\Omega\right] = \int_\Omega \bm{\bar{D}}\cdot\bm{\delta \bar{E}} d\Omega + \int_\Omega \Delta(\bm{D}\cdot\bm{\delta E}) d\Omega
\label{eq:LT3}
\end{equation}
\begin{equation}
\begin{split}
\int_\Omega \Delta(\bm{D}\cdot\bm{\delta E}) d\Omega &= \int_\Omega \bm{D}\cdot \Delta\bm{\delta E} d\Omega + \int_\Omega \Delta\bm{D} \cdot \bm{\delta E} d\Omega\\
&= \int_\Omega \delta \bm{E}\cdot\bm{\mu} \vdots\Delta\bm{\tilde{G}} d\Omega+\int_\Omega \bm{\delta E}\cdot \bm{e} : \Delta\bm{G}  d\Omega +\int_\Omega \bm{\delta E}\cdot \bm{\kappa} \cdot \Delta\bm{E}  d\Omega \\
\end{split}
\end{equation}\\
The linearization of the term $\int_{\Gamma} \bm{S^s} : \bm{\delta G^s} d\Gamma$ in Equation~\ref{eq:work1} is as follows,\\
\begin{equation}
L\left[\int_\Gamma \bm{S^s}:\bm{\delta G^s} d\Gamma\right] = \int_\Gamma \bm{\bar{S^s}}:\delta \bm{\bar{G^s}} d\Gamma + \int_\Gamma \Delta(\bm{S^s}:\bm{\delta G^s}) d\Gamma
\label{eq:LT4}
\end{equation}
\begin{equation}
\begin{split}
\int_\Gamma \Delta(\bm{S^s}:\bm{\delta G^s}) d\Gamma &= \int_\Gamma \bm{S^s}:\Delta(\bm{\delta G^s}) d\Gamma + \int_\Omega \bm{\delta G^s}:\Delta\bm{S^s} d\Gamma\\
&= \int_\Gamma \bm{S^s}:\Delta\bm{(\delta G^s)} d\Gamma + \int_\Gamma \bm{\delta G^s}:\bm{C^s}:\Delta\bm{ G^s} d\Gamma \\
&= \int_\Omega \bm{S^s}:\bm{P}\cdot[(\nabla_0 \bm{\delta u})^T (\nabla_0 (\Delta\bm{u}))]\cdot\bm{P} d\Gamma + \int_\Gamma \bm{\delta G^s}:\bm{C}:\Delta\bm{G^s} d\Gamma\\
\end{split}
\end{equation}\\
The algebraic forms of Equations~\ref{eq:LT1},\ref{eq:LT2},\ref{eq:LT3} and \ref{eq:LT4} are as follows,
\begin{equation}
\begin{split}
L\left[\int_\Omega \bm{S}:\bm{\delta G }d\Omega\right]  &= \bm{\delta u} \left(\int_{\Omega} \bm{B^T} \bm{\hat{R}} d\Omega\right)\;  + \bm{\delta u} \left(\int_\Omega \bm{B^T} \bm{C}\;\bm{B} d\Omega\right)  \bm{\Delta u}\\
&+ \bm{\delta u} \left(\int_\Omega \bm{B^T} \bm{e}\;\bm{B_\phi} d\Omega\right)  \bm{\Delta \phi} + \bm{\delta u} \left(\int_{\Omega} \bm{H_1^T} \bm{{R}}\bm{H_1}d\Omega\right)\;  \bm{\Delta u}
\end{split}
\label{Eq:ALT1}
\end{equation}\\
\begin{equation}
\begin{split}
L\left[\int_\Omega \bm{\tilde{S}}\vdots\bm{\delta \tilde{G}} d\Omega\right] &= \bm{\delta u} \left(\int_{\Omega} \bm{H_D^T} \bm{\hat{R}_D} d\Omega \right)\;+ \bm{\delta u} \left(\int_{\Omega} \bm{H_D^T} \bm{\mu^T} \bm{B_\phi} d\Omega \right)\;\bm{\Delta \phi} \\
&+  \bm{\delta u} \left(\int_{\Omega} \bm{H_1^T} \bm{R_D^T} \bm{H_2} d\Omega\right)\; \bm{\Delta u}
+  \bm{\delta u} \left(\int_{\Omega} \bm{H_2^T} \bm{R_D} \bm{H_1} d\Omega\right)\; \bm{\Delta u}
\end{split}
\label{Eq:ALT2}
\end{equation}
\begin{equation}
\begin{split}
L\left[\int_\Omega \bm{D}\cdot\bm{\delta E} d\Omega\right]& = -\bm{\delta \phi} \int_{\Omega} \bm{B_\phi}^T \bm{\hat{D}} d\Omega\;-\;\bm{\delta \phi} \left(\int_{\Omega} \bm{B_\phi}^T \bm{\mu}\bm{H_u}  d\Omega \right)\;\bm{\Delta u}\\
&-\;\bm{\delta \phi} \left(\int_{\Omega} \bm{B_\phi}^T \bm{e}\bm{B}  d\Omega \right)\;\bm{\Delta u}+ \bm{\delta \phi} \left(\int_{\Omega} \bm{B_\phi^T} \bm{\kappa} \bm{B_\phi} d\Omega \right)\;\bm{\Delta \phi}
\end{split}
\label{Eq:ALT3}
\end{equation}\\
\begin{equation}
\begin{split}
L\left[\int_\Gamma \bm{S^s}:\bm{\delta G^s}d\Gamma\right]  &= \bm{\delta u} \left(\int_{\Gamma} \bm{B^T} \bm{M_p}^T\bm{\hat{R}_s} d\Gamma\right)\;  + \bm{\delta u} \left(\int_\Gamma \bm{B^T}\bm{M_p}^T \bm{C^s}\;\bm{M_p}\;\bm{B} d\Omega\right)  \bm{\Delta u}\\
&\bm{\delta u} \left(\int_{\Gamma} \bm{H_1^T} \bm{P_n^T}\bm{{R_s}}\bm{P_n}\bm{H_1}d\Gamma\right)\;  \bm{\Delta u}
\end{split}
\label{Eq:ALT4}
\end{equation}\\
All the matrices involved in Equations~\ref{Eq:ALT1}, \ref{Eq:ALT2}, \ref{Eq:ALT3}, \ref{Eq:ALT4} are presented in \ref{app2}. The final algebraic form of linearization of Equation~\ref{eq:work1} can be written as,
\begin{equation}
\bm{K}\bm{\Delta U}=\bm{F^{ext}}-\bm{F^{int}}
\label{Eq:NLEq}
\end{equation}
where,
\begin{equation}
\begin{split}
&\bm{K}= \left[\begin{array}{c c}
\bm{K_{uu}} & \bm{K_{u\phi}}\\ 
\bm{K_{\phi u}} & \bm{K_{\phi\phi}}\end{array}\right] \\
& \bm{\Delta U}= \left[\begin{array}{c}
\bm{\Delta u} \\
\bm{\Delta\phi}
\end{array}\right]
\end{split}
\end{equation}\\
\begin{equation}
\begin{split}
\bm{K_{uu}} &= \int_{\Omega} \bm{B^T} \bm{C} \bm{B} d\Omega + \int_{\Omega} \bm{H_1^T} \bm{R} \bm{H_1} d\Omega \\&
+ \int_{\Omega} \bm{H_1^T} \bm{R_D^T} \bm{H_2} d\Omega + \int_{\Omega} \bm{H_2^T} \bm{R_D} \bm{H_1} d\Omega\\
&+ \int_{\Gamma} \bm{B^T} \bm{{M_p}^T} \bm{C^s} \bm{{M_p}} \bm{B} d\Gamma + \int_{\Gamma} \bm{H_1^T}\bm{P_n} \bm{R_s}\bm{P_n} \bm{H_1} d\Gamma
\end{split}
\end{equation}
\begin{equation}
\bm{K_{u\phi}} = \int_{\Omega} \bm{B^T} \bm{e^T} \bm{B_\phi} d\Omega  +\int_{\Omega} \bm{H_D^T} \bm{\mu^T} \bm{B_\phi} d\Omega  = \bm{K_{\phi u}}^T
\end{equation}
\begin{equation}
\bm{K_{\phi\phi}} = -\int_{\Omega} \bm{B_\phi^T} \bm{\kappa} \bm{B_\phi} d\Omega 
\end{equation}
\begin{equation}
\begin{split}
& \bm{F^{int}} = \left[\begin{array}{c}
\bm{F_u} \\ \bm{F_\phi}
\end{array}\right] \\
& \bm{F_u} =  \int_{\Omega} \bm{B^T} \bm{\hat{R}} d\Omega +  \int_{\Omega} \bm{H_D^T} \bm{\hat{R}_D} d\Omega  + \int_{\Gamma} \bm{B^T} \bm{M_p}^T\bm{\hat{R}_s} d\Gamma\;\\
& \bm{F_\phi} = \int_{\Omega} \bm{B_\phi}^T \bm{\hat{D}} d\Omega
\end{split}
\end{equation}
The nonlinear equation~\ref{Eq:NLEq} is solved by using the Newton-Raphson iterative scheme. Solving this equation, gives the deflection and voltage responses of flexoelectric structures that undergo large deformations.
\section{Numerical Examples}
\label{sec5}
In this section, the proposed meshfree formulation is utilised to analyse flexoelectric cantilever beams accounting for surface effects and also to study the influence of geometric nonlinearity on their voltage output. As an initial step, the meshfree formulation is validated by determining the energy conversion factor (ECF) of a cantilever beam. The ECF is given by the ratio between stored electrical energy and mechanical energy in the flexoelectric structure.
\begin{table}
\caption{Electromechanical properties of STO} 
\centering  
\begin{tabular}{c c c c} 
\hline
Elastic Constants & Dielectric constants & Flexoelectric constants \cite{margaranti}\\ [0.5ex] 
\hline                  
\\
$C_{11}$=310  $GPa$  & $\kappa_{11}$=2.66 $C/(GV-m)$ & $\mu_{11}$=-0.26 $nC/m$ \\ 
$C_{12}$=115 $GPa$  &$\kappa_{33}$=2.66 $C/(GV-m)$  & $\mu_{12}$=-3.74 $nC/m$  \\
$C_{22}$=310   $GPa$  & &$\mu_{44}$=-3.56 $nC/m$         \\
$C_{66}$=54    $GPa$       &           &           \\    
\hline 
\end{tabular}
\label{table:STO}
\end{table}

\subsection{Validation: ECF}
\begin{figure} \begin{center} 
{\includegraphics[scale=0.65]{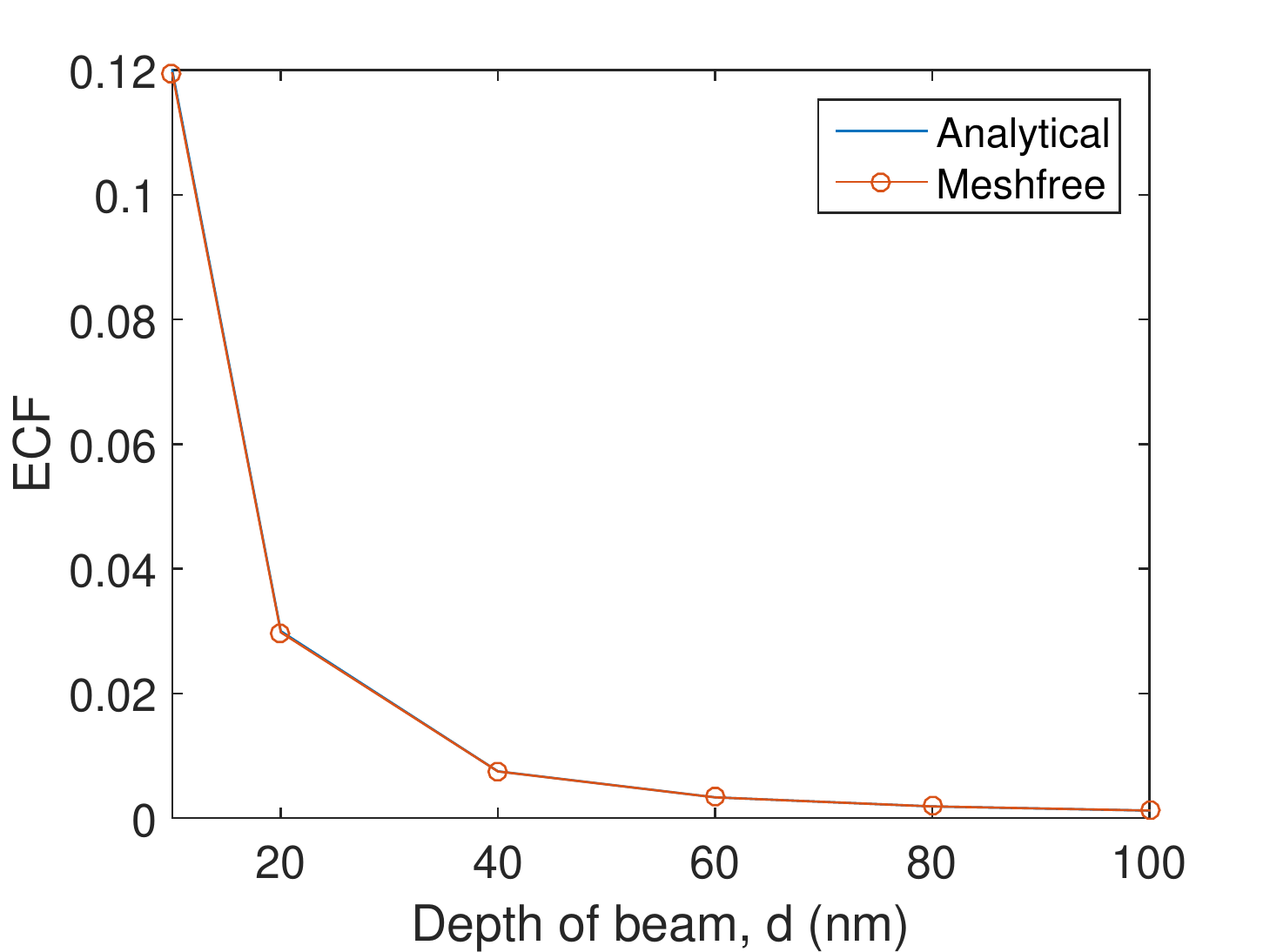}}
\caption{The variation of ECF, $k^2$ with beam depth for a one dimensional beam model.}
\label{fig:valid} \end{center} \end{figure}

A cantilever beam subjected to a mechanical point load at the free end is analysed in order to validate the proposed meshfree formulation. The cantilever beam has an aspect ratio of 6. The Young's modulus, Y is assumed to be 100 GPa. For validation, only the flexoelectric constant $\mu_{12}$ and dielectric constant $\kappa_{22}$ are considered non-zero and they are assumed to be 10 $nC/m$ and 1 $nC/Vm$ respectively. The beam is discretized by 121 $\times$ 21 nodes with uniform spacing and the background mesh for numerical integration is 120 $\times$ 20. The variation of the energy conversion factor (ECF) with decreasing depth of the 1D flexoelectric beam is shown in Figure~\ref{fig:valid}. The figure shows good agreement between the numerical and analytical $k^2$ values. The analytical energy conversion factor, $k^2$ for a 1D model is given in \cite{majdoubPRB2008} as, \\
\begin{equation}
ECF^{anl} = k = \sqrt{\frac{12}{\kappa Y}\left(\frac{\mu}{d}\right)^2}
\end{equation}
\subsection{Validation: Tube model}
\label{2D-ex}
\begin{figure} \begin{center} 
\subfigure[]{{\includegraphics[scale=0.45]{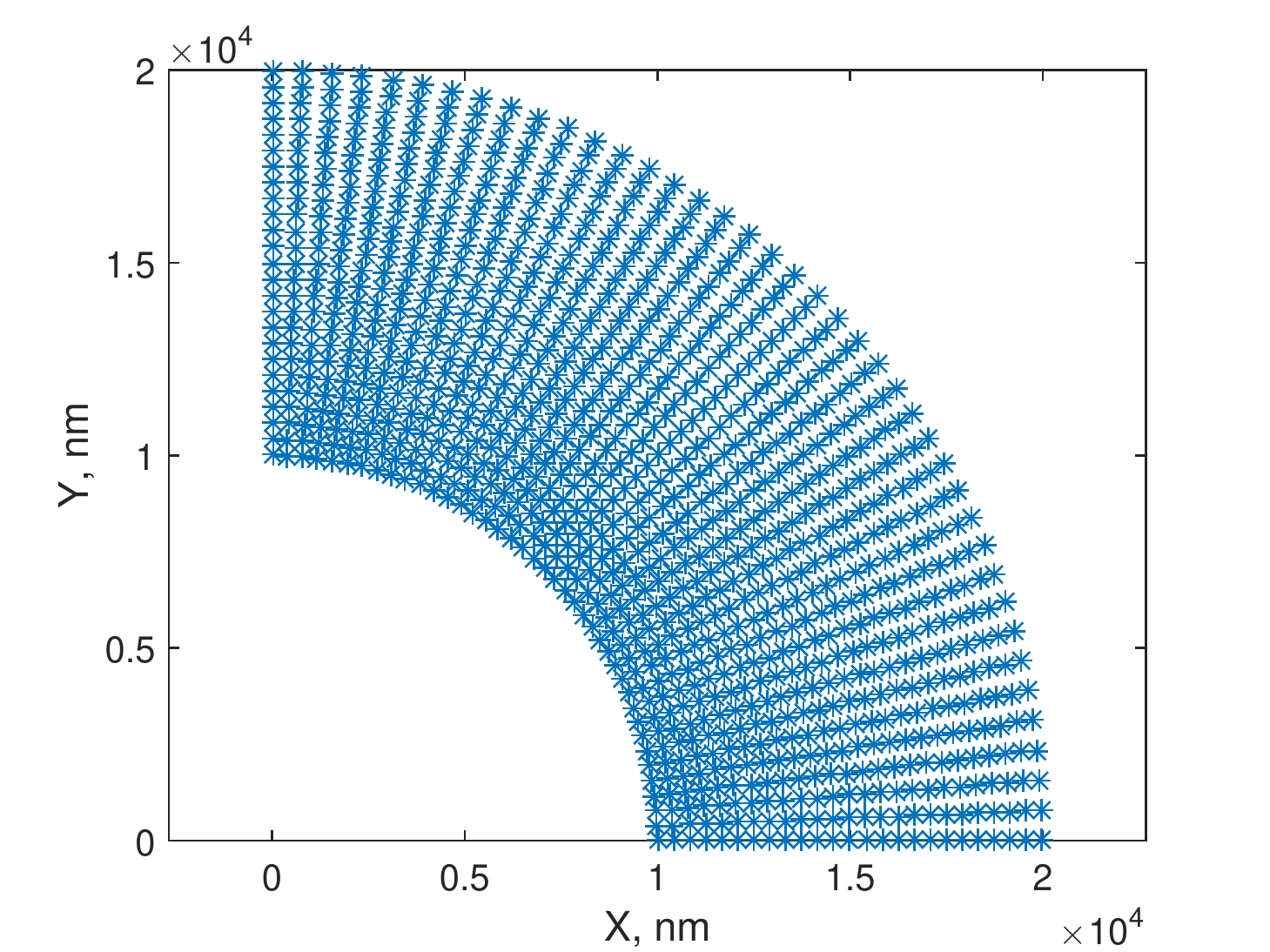}}\label{fig:tube}}
\subfigure[]{{\includegraphics[scale=0.45]{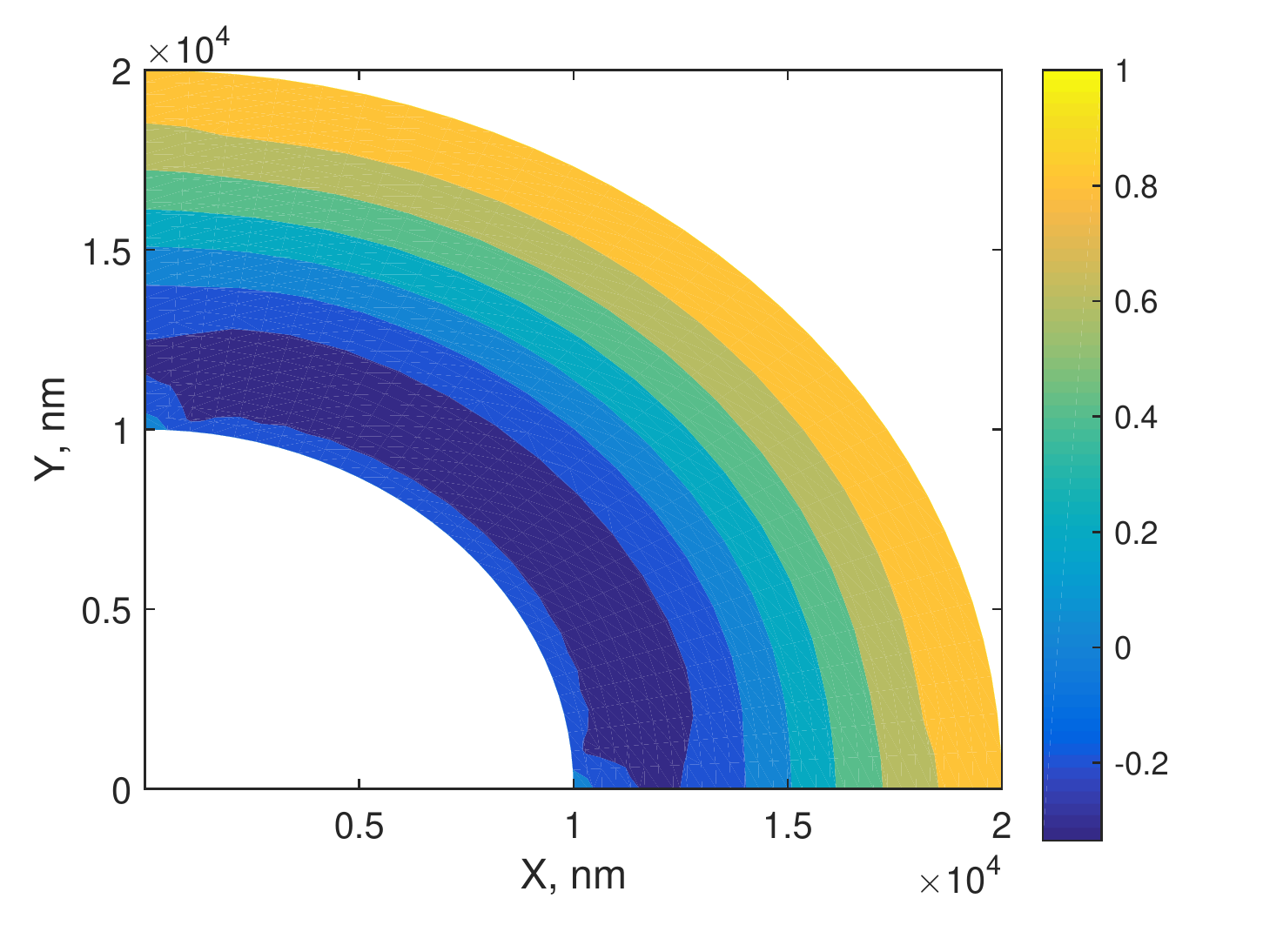}}\label{fig:pot_tube}}
\subfigure[]{{\includegraphics[scale=0.45]{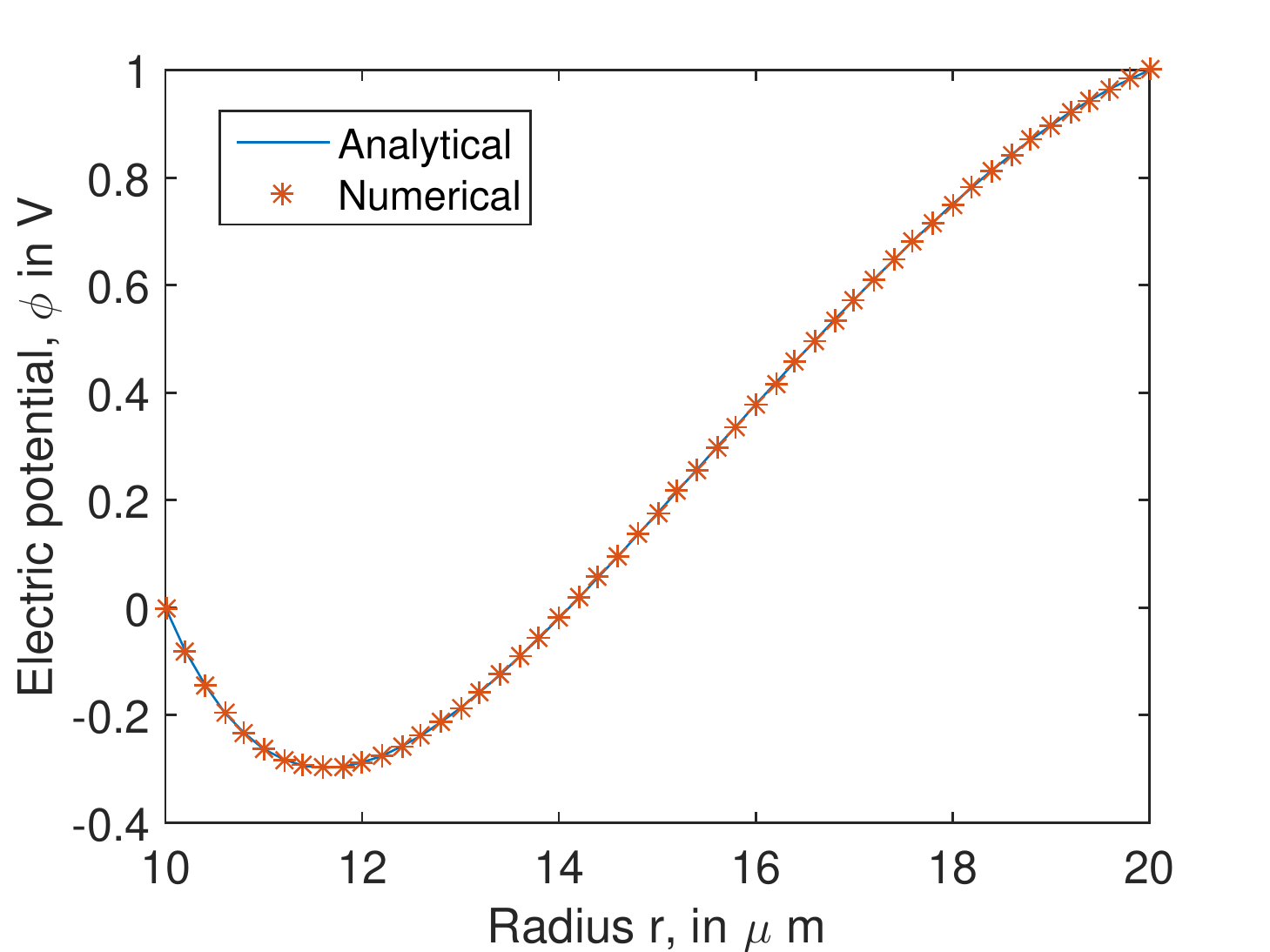}}\label{fig:pot_l_tube}}
\caption{(a) Nodal distribution for the quarter tube model, (b) Electric potential distribution across the tube cross section, (c) The variation in output voltage along the radius of the tube.}
\end{center} \end{figure}

In order to further validate the proposed formulation, a  flexoelectric tube made of STO is analysed with plane strain assumption. The tube with an inner radius, $r_i$ of 10 $\mu m$ and outer radius, $r_o$ of 20 $\mu m$  is subjected to a radial displacement of 0.045 $\mu m$ and 0.05 $\mu m$ at $r_i$ and $r_o$ respectively \cite{Mao_FE,Shen_FE}. The tube is grounded along the inner face and a voltage of 1 $V$ is applied along the outer face. The nodal distribution of the quarter model is as shown in Figure~\ref{fig:tube}. The distribution of electric potential obtained for a length scale, $l_0$ of 2 $\mu m$ is shown in figure~\ref{fig:pot_tube}. The variation of electric potential along the thickness of the tube is shown in figure~\ref{fig:pot_l_tube}. The analytical results for the flexoelectric tube model with assumed material parameters is derived by Mao \textit{et al}.~\cite{Mao_FE}. The results presented in Figure~\ref{fig:pot_l_tube} shows good agreement between the analytical 
 and numerical results.
\\
\begin{figure} \begin{center} 
{\includegraphics[scale=0.5]{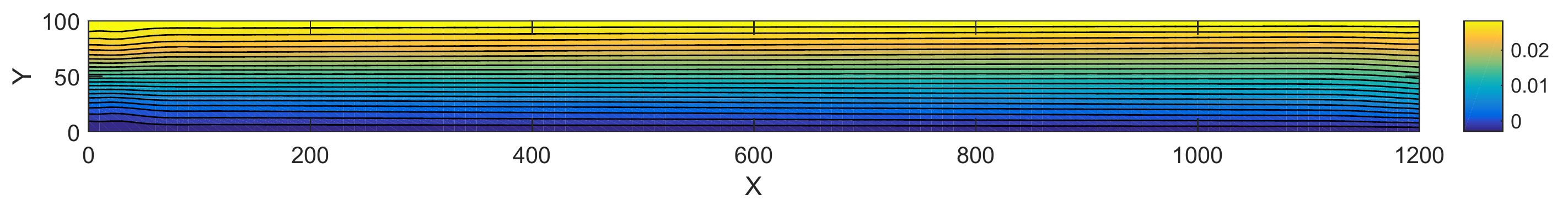}}
\caption{The potential distribution in a flexoelectric beam made of STO.}
\label{fig:pot_STO} \end{center} \end{figure}
\begin{figure} \begin{center} 
{\includegraphics[scale=0.65]{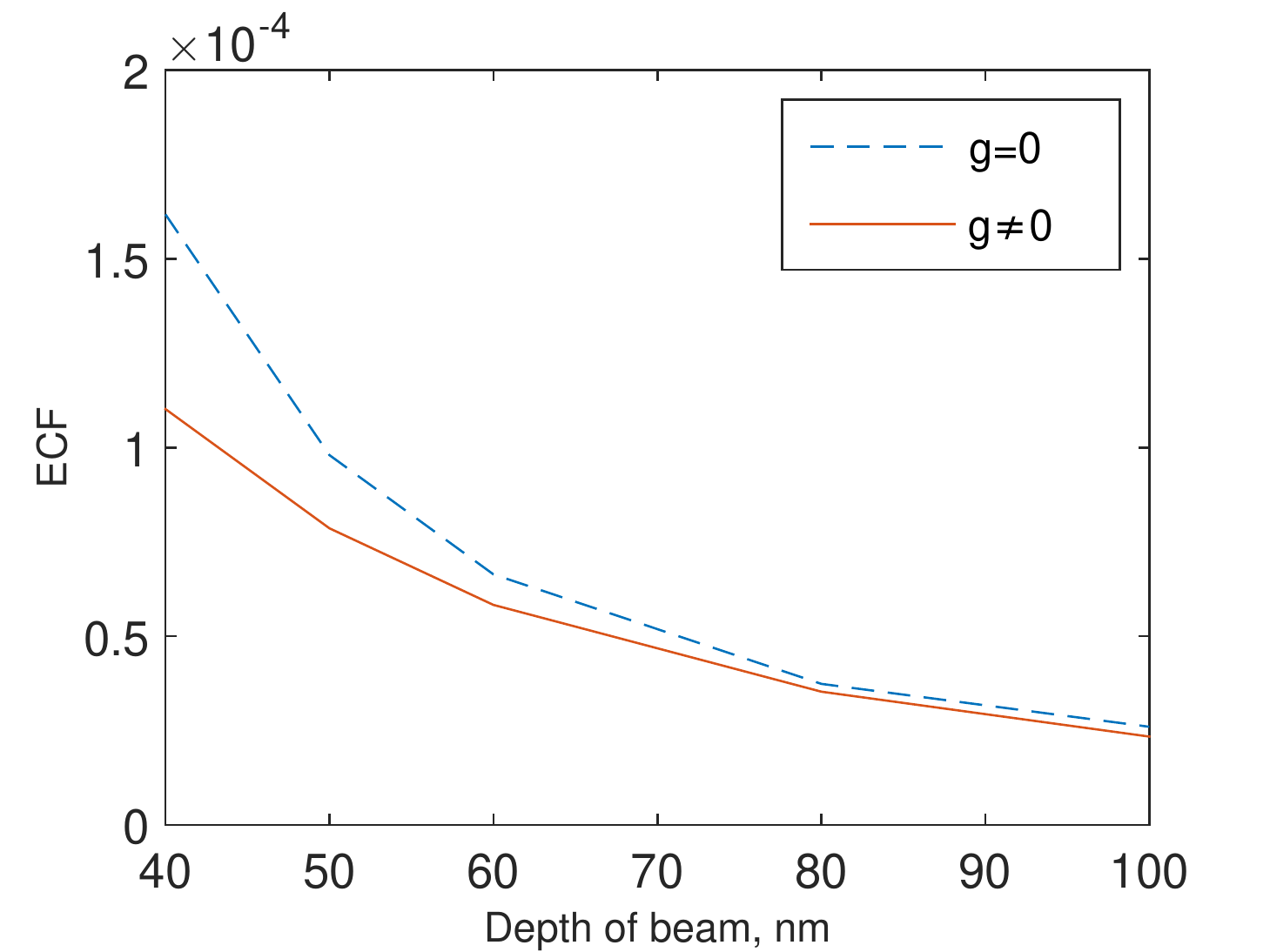}}
\caption{The variation of ECF with depth of beam including and excluding strain gradient tensor, $g$. The length scale is taken as 5 nm.}
\label{fig:2D-ECF} \end{center} \end{figure}
\subsection{Pure flexoelectricity}
\begin{figure} \begin{center} 
{\includegraphics[scale=0.65]{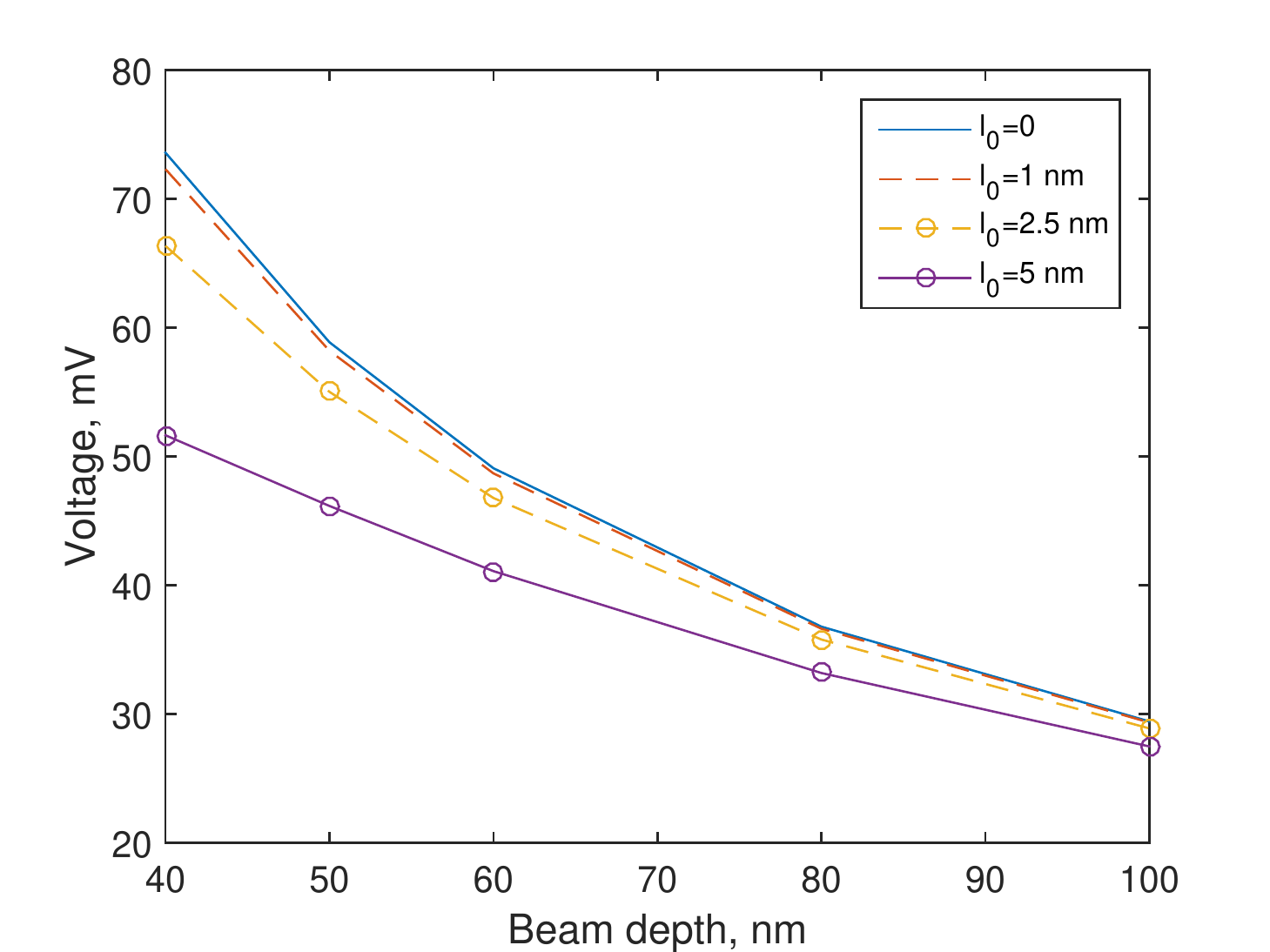}}
\caption{Output voltage for varying beam depths for internal length scales of 0,1,2.5 and 5 nm.}
\label{fig:V_g} \end{center} \end{figure}

A cantilever beam made of cubic STO is analysed in this section. The material properties of STO is given in Table~\ref{table:STO}. The bottom and top face of the beam are coated with electrodes. The bottom electrode is grounded and the top electrode is free to have a potential value.\\
The dimension of the beam is 1200 $\times$ 100 nm. The beam is subjected to a point load of 10 nN at the free end. The length scale, $l_0$ of $\bm{g}$ tensor is taken as 0 (i.e.) $\bm{g}$ is not considered in this analysis. The  beam has an almost linear variation of potential along the beam depth as shown in Figure~\ref{fig:pot_STO}. The potential obtained at the top face is 30 mV. Now if we fix the aspect ratio to be 12, and reduce the beam depth for instance to 40 nm, then the potential at the top face is 74 mV. If the tensor, $\bm{g}$ is included in the analysis with a length scale, $l_0$ of 5 nm, then the potential obtained at the top face of the 40 nm beam depth reduces to 52 mV.\\
The change in energy conversion factor with depth of beam for an aspect ratio of 12, excluding and including $\bm{g}$ tensor is shown in Figure~\ref{fig:2D-ECF}. The energy conversion factor for 40 nm beam depth with and without including $\bm{g}$ tensor are 1.1e-4 and 1.61e-4 respectively. The inclusion of strain gradient elasticity increases the stiffness of the beam, reduces the voltage obtained and as a result reduces the ECFs. Note that the difference between the energy conversion factors with and without the $\bm{g}$ tensor increases with reduction in depth of the beam.\\
On the other hand, the influence of the length scale, $l_0$ on the output voltage is shown in Figure~\ref{fig:V_g}. It can be seen that, when the length scale is increased from 0 to 5 nm, the output voltage of the 100 nm beam reduces from 30 mV to 27.5 mV (6.3$\%$ reduction). While for the same increase in length scale, the output voltage of the 40 nm beam reduces from 73.6 mV to 51.7 mV (28.6$\%$ reduction).\\
\\

\subsection{Flexoelectricity and surface effects}
In this section, we analyse a Zinc Oxide nano cantilever beam. The interplay between piezoelectric, surface elastic, surface piezoelectric and flexoelectric effects is studied. Zinc Oxide is the ideal material for performing this study because it is widely used in several nanoscale energy harvesters~\cite{zhu,wang_zl} and studies on surface properties of Zinc Oxide~\cite{Yvonnet} is available.

The cantilever ZnO beam is of length, 120 nm and width, 15 nm. A point load of 10 nN is applied in x-direction at the mid-point of the top face. The beam is fixed at the bottom and free at the top. The beam is poled along the length (y-direction). The bottom end of the beam is grounded. The elastic, piezoelectric, surface elastic and surface piezoelectric properties of ZnO are given in Table~\ref{table:bulk} and Table~\ref{table:surf}. The residual stress, $\bm{\tau^s}$ and residual electric displacement, $\bm{\omega^s}$ are not considered in the study. The flexoelectric constant of ZnO, $\mu_{11}$, $\mu_{12}$ and $\mu_{44}$ are assumed to be 2 nC/m, 2 nC/m and 0.5 nC/m respectively.\\
\begin{table}
\caption{Material properties of bulk ZnO} 
\centering  
\begin{tabular}{c c c} 
\hline
Elastic Constants & Piezoelectric constants & Dielectric constants\\ [0.5ex] 
\hline                  
\\
$C_{11}$=206  $GPa$ & $e_{31}$=-0.58 $C/m^2$ & $\kappa_{11}$=0.0811 $C/(GV-m)$   \\ 
$C_{12}$=117 $GPa$ & $e_{33}$=1.55 $C/m^2$ &$\kappa_{33}$=0.112 $C/(GV-m)$     \\
$C_{22}$=211  $GPa$ & $e_{15}$=-0.48   $C/m^2$ &           \\
$C_{66}$=44.3   $GPa$ &                       &           \\    
\hline 
\end{tabular}
\label{table:bulk}
\\
\caption{Material properties of ZnO surface} 
\centering  
\begin{tabular}{c c c} 
\hline
Elastic Constants & Piezoelectric constants  \\ [0.5ex] 
\hline                  
\\
$C_{11}^s$=44.2  $N/m$ & $e_{31}^s$=-0.216 $nC/m$ \\ 
$C_{12}^s$=14.2 $N/m$ & $e_{33}^s$=0.451 $nC/m$  \\
$C_{22}^s$=35   $N/m$ & $e_{15}^s$=-0.253   $nC/m$           \\
$C_{66}^s$=11.7   $N/m$ &                                 \\    
\hline 
\end{tabular}
\label{table:surf} 
\end{table}

\begin{figure} \begin{center} 
\subfigure[]{{\includegraphics[scale=0.32]{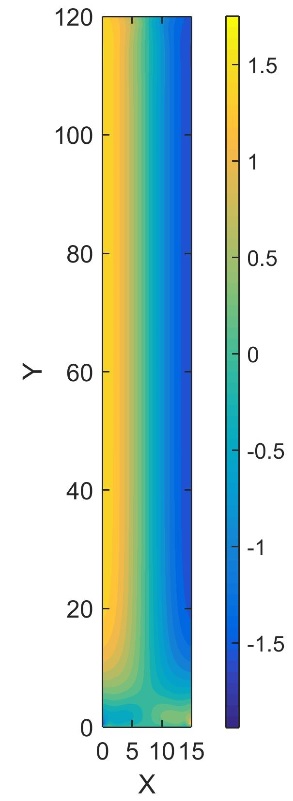}}\label{fig:pot_Zno1}}
\subfigure[]{{\includegraphics[scale=0.315]{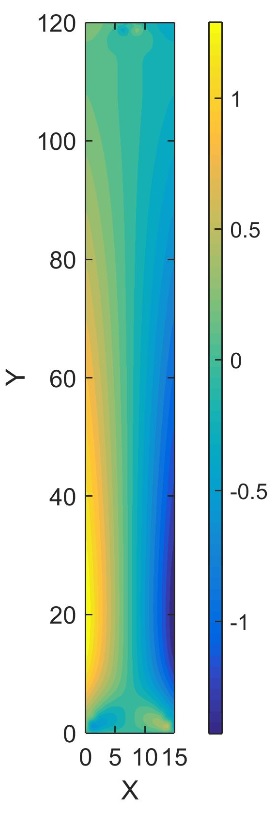}}\label{fig:pot_Zno2}}
\caption{The potential distribution in the ZnO beam considering (a) Flexoelectricity, piezoelectricity and surface effects, (b) Pure Flexoelectricity.}
\label{fig:pot_Zno} \end{center} \end{figure}

\begin{figure} \begin{center} 
{\includegraphics[scale=0.65]{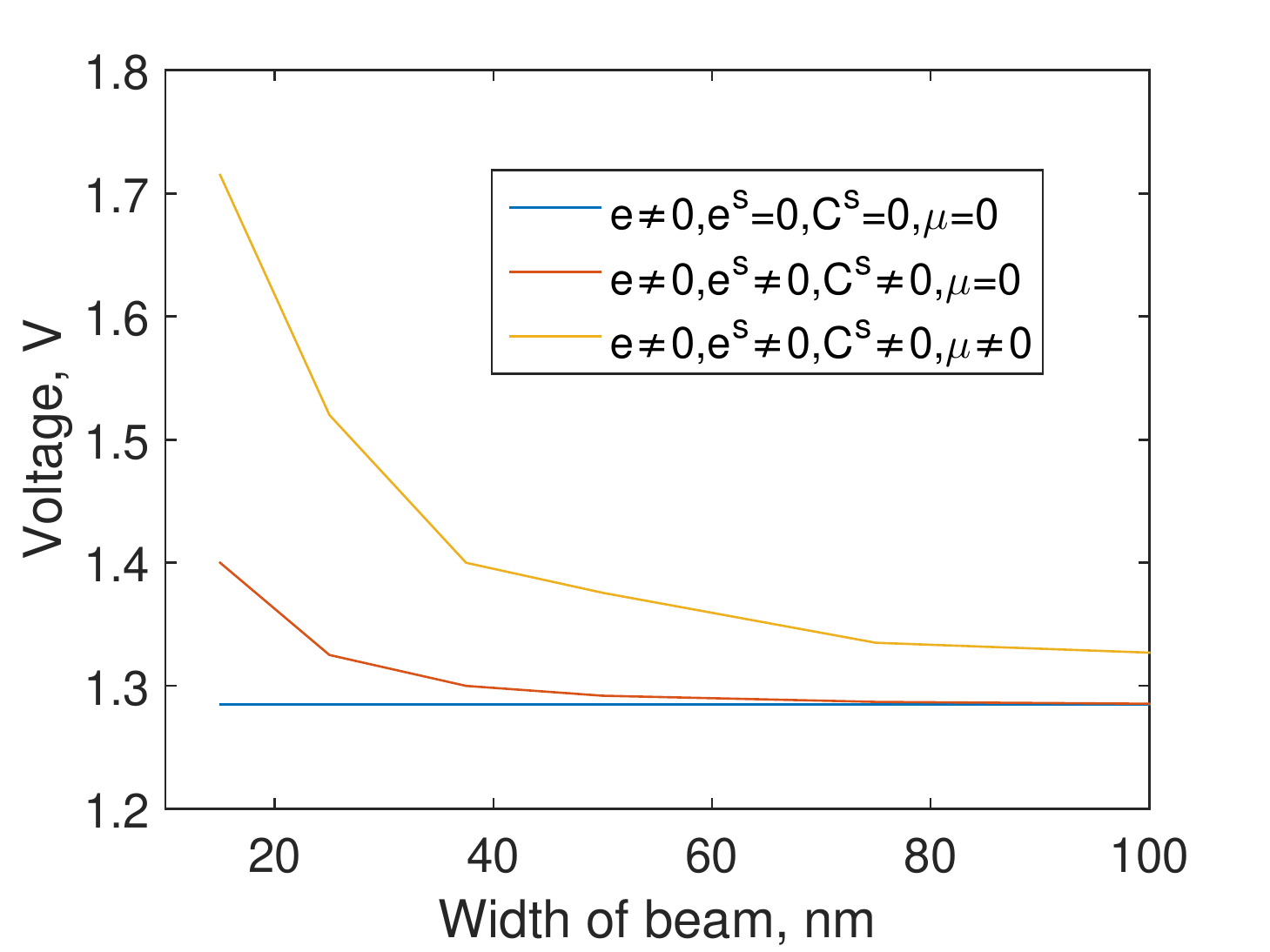}}
\caption{The variation of output voltage for ZnO beam with width for three cases, Piezoelectricity, Piezoelectricity+surface effects and Piezoelectricity+Surface effects+Flexoelectricity.}
\label{fig:comp_Zno} \end{center} \end{figure}

The potential distribution across the beam width is shown in Figure~\ref{fig:pot_Zno}. The combination of bulk piezoelectricity and surface elastic effect results in a potential of +1.18 V to -1.18 V at the top face. The combination of bulk piezoelectricity and surface piezoelectric effect leads to a potential varying from +1.5 V to -1.5 V at the top face of the beam.  Finally, the combination of bulk piezoelectricity, bulk flexoelectricity, surface elasticity and surface piezoelectricity results in a potential of +1.7 V to -1.7 V at the top face as shown in Figure~\ref{fig:pot_Zno1}. The variation of electric potential considering only flexoelectric effect is +0.3 V to -0.3 V at the top face (Figure~\ref{fig:pot_Zno2}).
The relative influence of the different phenomenon on the output voltage is shown in Figure~\ref{fig:comp_Zno}. The contribution of flexoelectric effect to output voltage is higher compared to the contribution of surface effects and the difference in the contributions to output voltage increases as the beam width decreases. For a 40 nm wide beam, the flexoelectric and surface effect contributions are 7$\%$ and 1$\%$ respectively. While for a width of 15 nm, the difference between the contributions is higher, the flexoelectric and surface effect contributions are 18$\%$ and 7$\%$ respectively.

\begin{figure} \begin{center} 
{\includegraphics[scale=0.65]{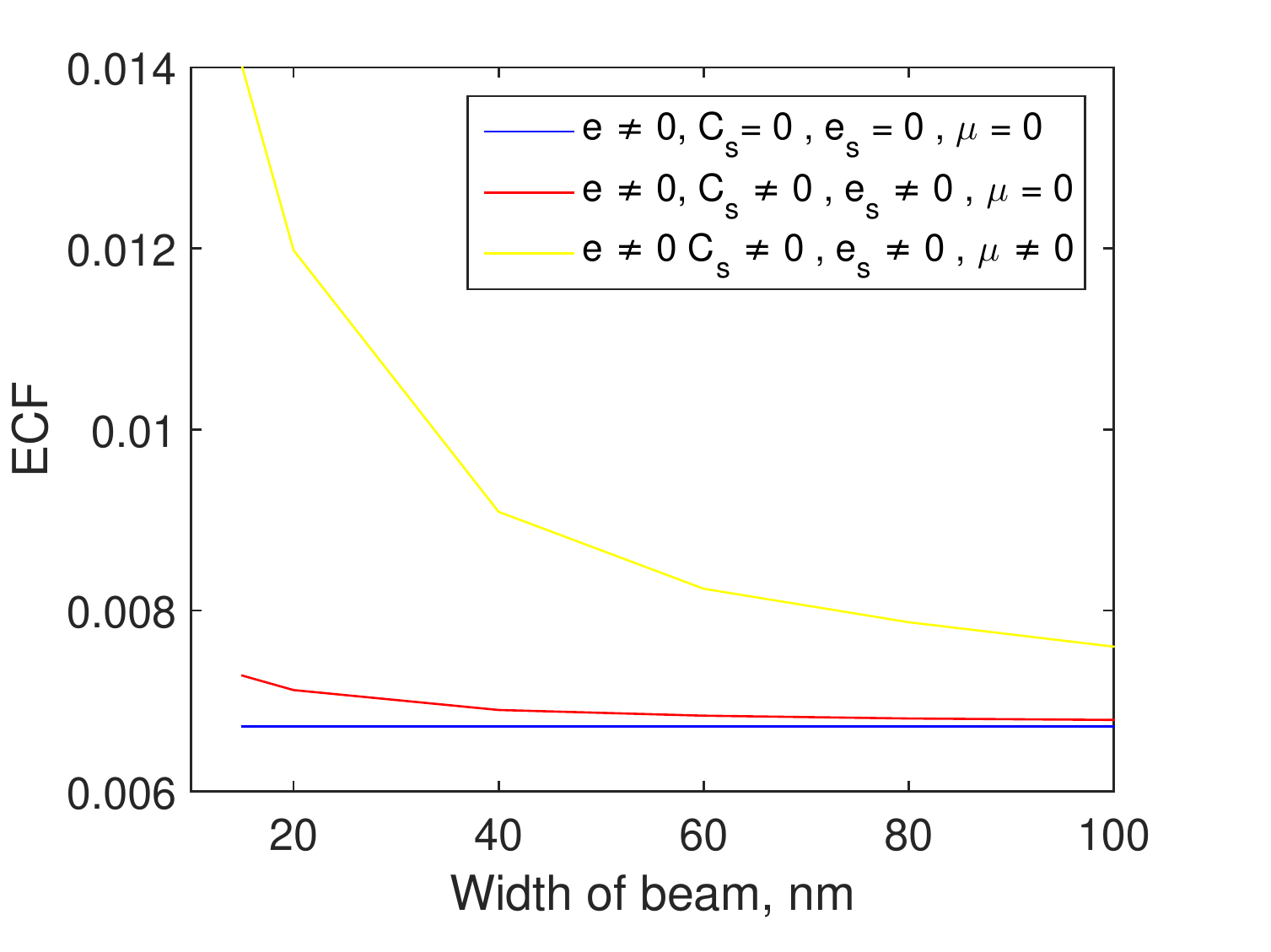}}
\caption{The variation of ECF for ZnO beam with width for three cases, Piezoelectricity, Piezoelectricity+Surface effects and Piezoelectricity+Surface effects+Flexoelectricity.}
\label{fig:ECF_surf_flexo} \end{center} \end{figure}

The change in ECF with width of beam is shown in Figure~\ref{fig:ECF_surf_flexo}. The pattern is similar to the one obtained for output voltage. The ECF for pure Piezoelectricity and Piezo$+$Surface effects for 15 nm wide beam are 0.0067 and 0.00728 respectively.  While for the same beam width, the ECF considering Piezo$+$Surface effects$+$Flexoelectricity is 0.014. The percentage contribution of flexoelectricity and surface effects to the total ECF are 48$\%$ and 8$\%$ respectively. There is a discrepancy in the percentage contribution of flexoelectricity to ECF and output voltage. This is because the potential due to flexoelectricity reaches its peak near the fixed end and reduces significantly along the length. So, though the flexoelectric contribution to total ECF is 48$\%$, the contribution of flexoelectricity to total voltage measured at the top face (y=120 nm) is only 18$\%$.\\
\subsection{Flexoelectricity and surface effects: Geometric nonlinearity}
\label{sec:5.4}
In this section, the flexoelectric response in the nonlinear regime is studied. The nonlinearity emerges due to large deformation of the flexoelectric cantilever beam. The flexoelectric beam is assumed to be made of STO, in addition to flexoelectricity, the surface elasticity of STO is also considered. The material properties of STO are given in Table~\ref{table:STO}. The surface elastic constants of STO are not available in literature and are assumed to be, $C^s_{11} = C^s_{22} = 310~GPa$, $C^s_{12} = 115~GPa$ and $C^s_{66} = 54~GPa$. The beam is subjected to mechanical deformation by a point load of 10 nN at the free end. The load is applied in increments of 1 nN. In each increment, the tangent stiffness matrix is determined and the Newton-Raphson method is adopted to minimize the residual. Two different beams each of thickness 50 nm and 30 nm respectively are analysed. The aspect ratio of the beams is fixed as 12. The fixed end of the cantilever is grounded.

\begin{figure} \begin{center} 
\subfigure[]{{\includegraphics[scale=0.55]{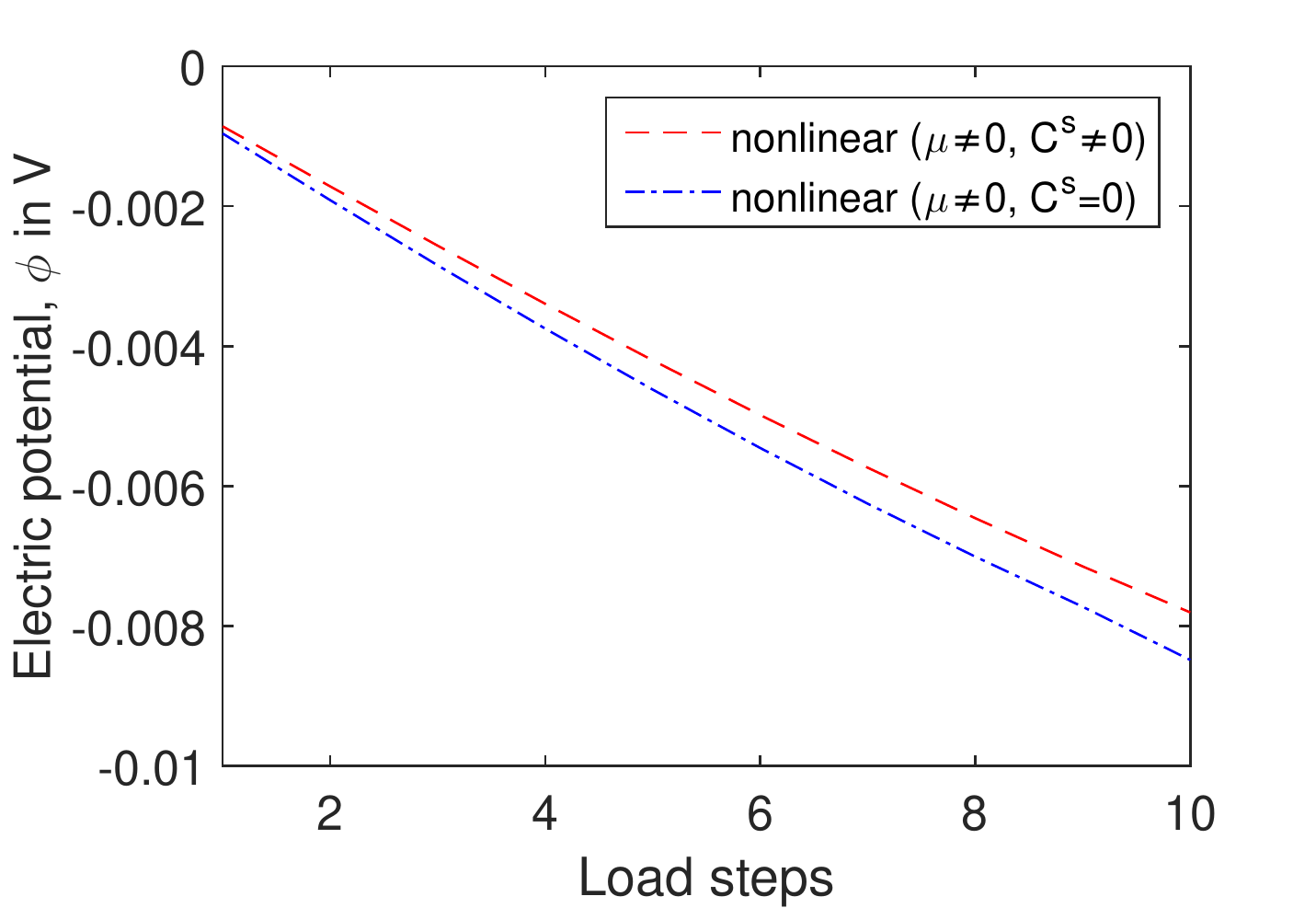}}\label{fig:nonlinear1}}
\subfigure[]{{\includegraphics[scale=0.5]{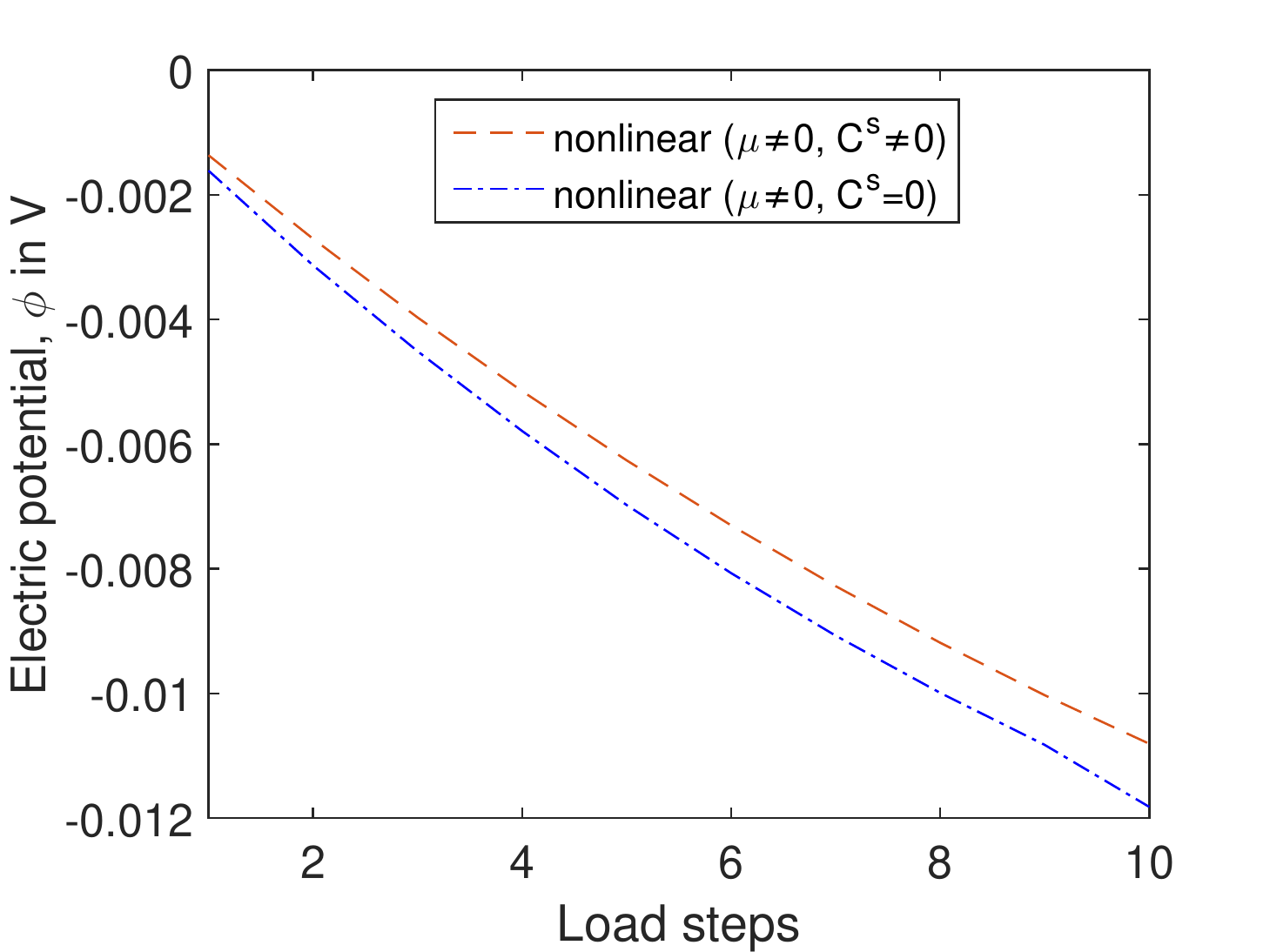}}\label{fig:nonlinear2}}
\caption{The variation in voltage with load increments for flexoelectric beam made of STO of depth (a) 50 nm and (b) 30 nm (Aspect ratio=12).}
\label{fig:nonlinear} \end{center} \end{figure}

The variation of maximum output voltage with load steps is shown in Figure~\ref{fig:nonlinear}. As shown in Figure~\ref{fig:nonlinear1}, at the end of ten load steps the voltage due to flexoelectricity and surface elasticity for 50 nm thick beam is 7.8 mV. The ratio between nonlinear and linear voltage is 0.9 (i.e.) the nonlinear voltage deviates from the linear response by $10\%$. If surface effects are not considered, then the final output voltage is 8.5 mV. In case of 30 nm thick beam, as shown in Figure~\ref{fig:nonlinear2}, the final output flexoelectric voltage is 11 mV and 12 mV considering and ignoring surface effects respectively.

\begin{figure} \begin{center} 
\subfigure[]{{\includegraphics[scale=0.5]{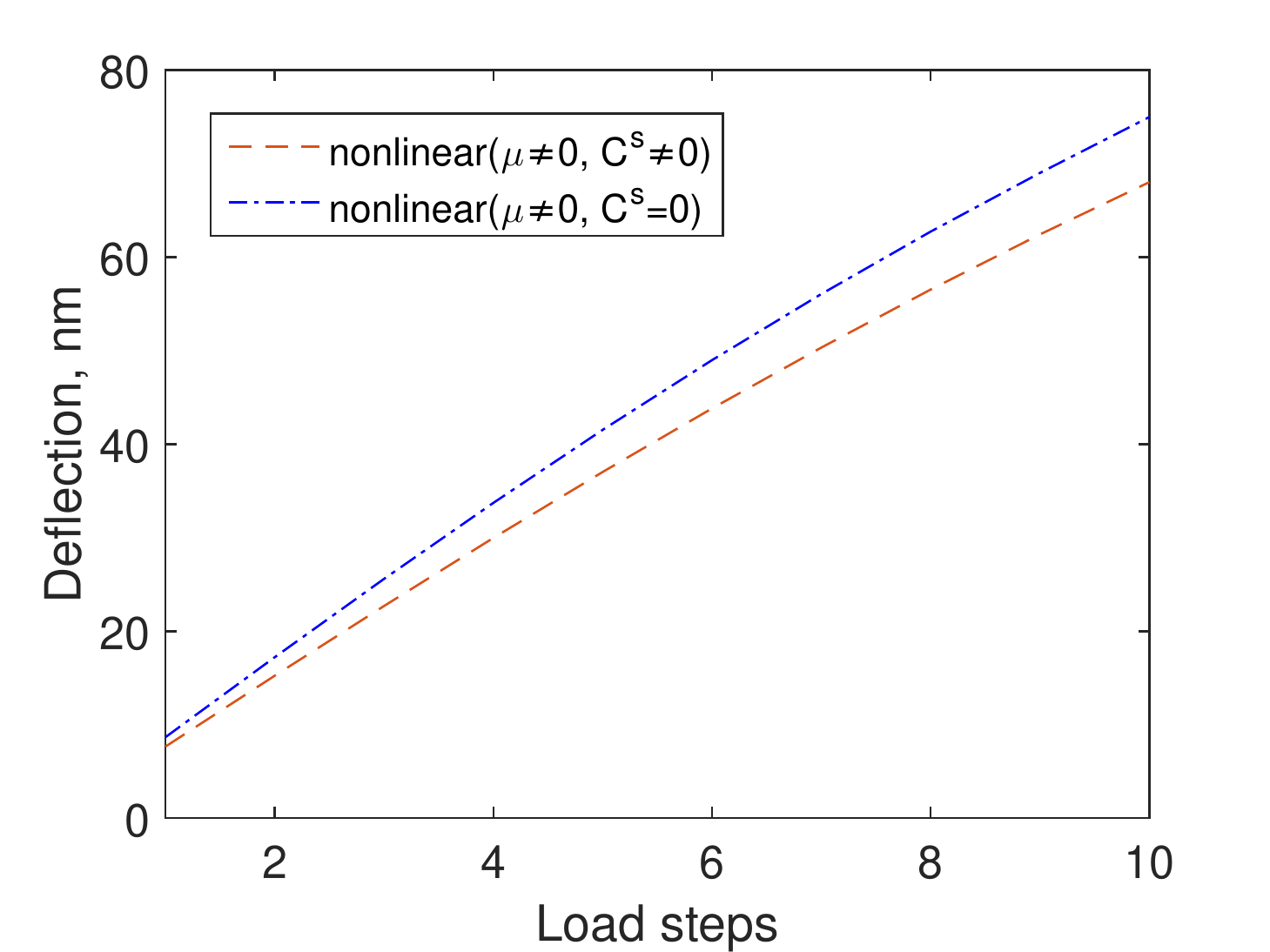}}\label{fig:F_U1}}
\subfigure[]{{\includegraphics[scale=0.5]{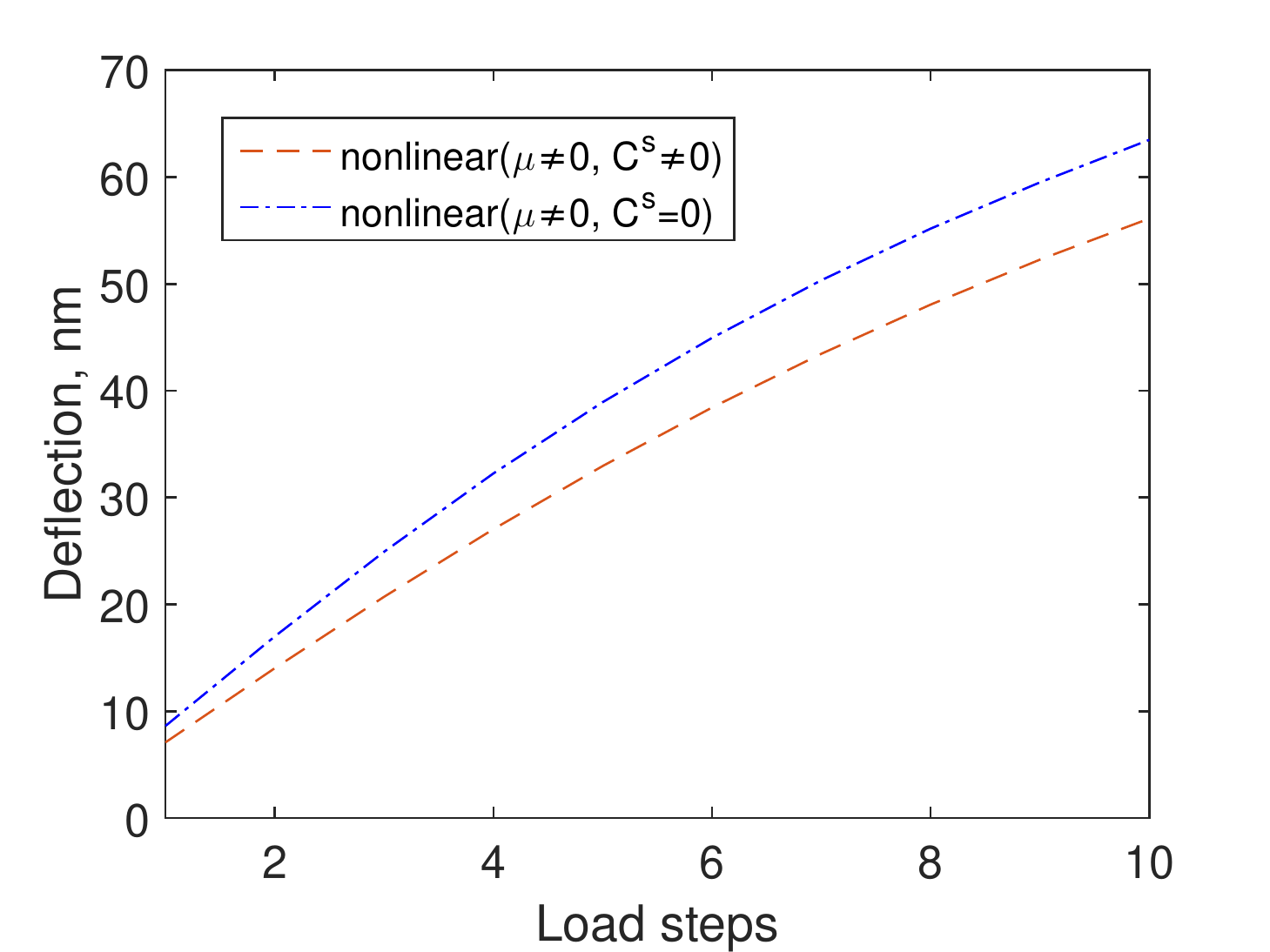}}\label{fig:F_U2}}
\caption{The load deflection curve for flexoelectric beam made of STO of depth (a) 50 nm and (b) 30 nm (Aspect ratio=12).}
\label{fig:F_U} \end{center} \end{figure}

\begin{figure} \begin{center} 
{\includegraphics[scale=0.65]{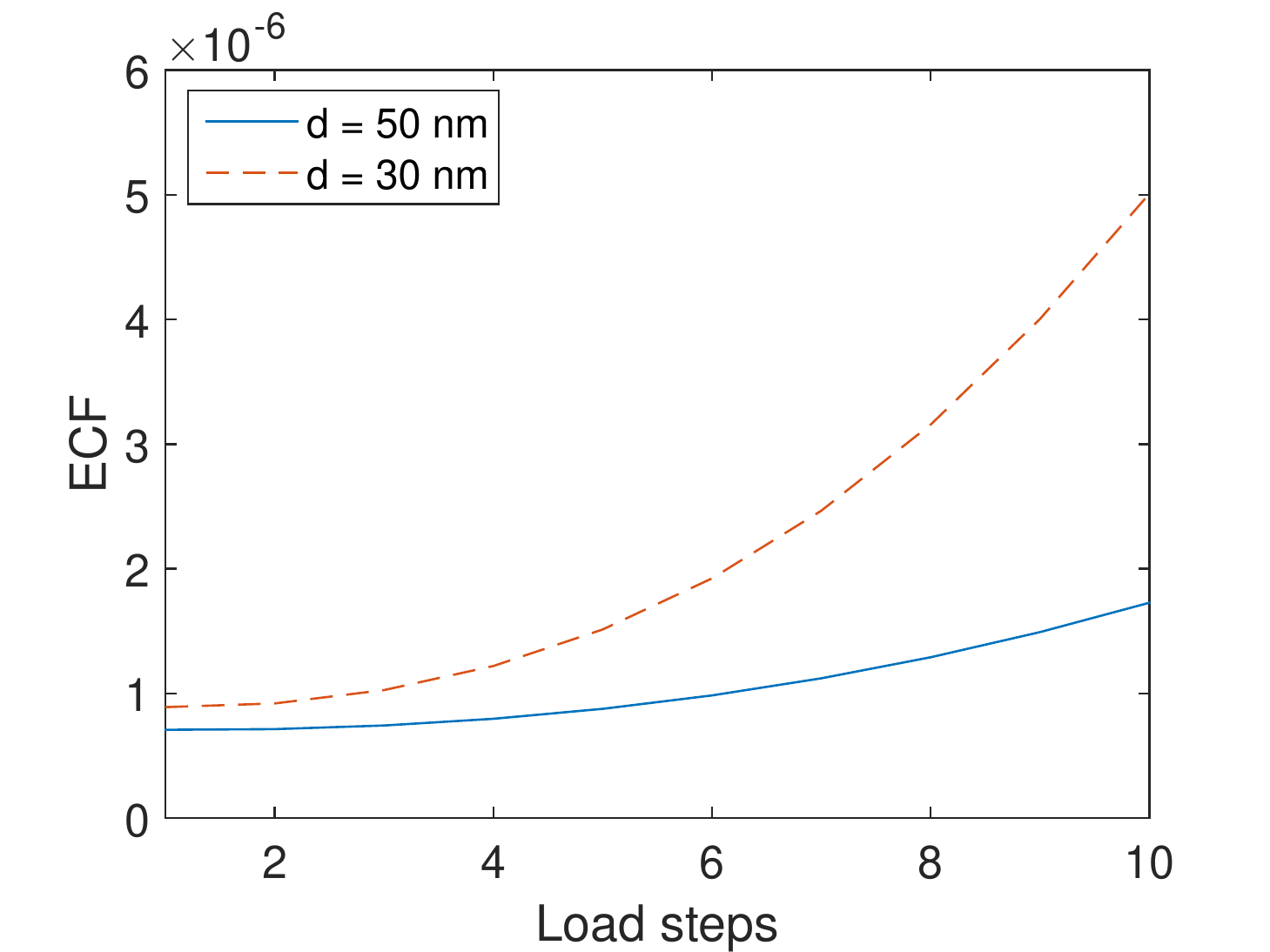}}
\caption{The variation in ECF with load increments for flexoelectric beam made of STO of depth 50 nm and 30 nm (Aspect ratio =12).}
\label{fig:nonlin_ECF} \end{center} \end{figure}

The variation of free end deflection with load steps is shown in Figure~\ref{fig:F_U}. The free end deflection of 50 nm thick beam after ten load steps is 68 nm (Figure~\ref{fig:F_U1}). The ratio between nonlinear and linear deflection for a load value of 10nN for 50 nm thick beam is 0.89 (i.e.) a deviation of nonlinear displacement is $11\%$ from linear displacement. In the absence of surface effects, the final output deflection of 50 nm thick beam is 75 nm. For the case of 30 nm thick beam, as shown in Figure~\ref{fig:F_U2}, the final deflection is 56 nm and 63.5 nm considering and ignoring surface effects respectively.From Figures~\ref{fig:nonlinear} and~\ref{fig:F_U}, it is clear that the beam becomes stiffer in the presence of surface elasticity, which in turn leads to reduction in output voltage. In case of 50 nm thick beam, due to surface elasticity the absolute value of final voltage reduces from 8.5 mV to 7.8 mV. While for 30 nm thick beam, the voltage reduction is from 12 m
 V to 11 mV.  The variation of energy conversion factor with load steps is shown in Figure~\ref{fig:nonlin_ECF}. The rate of increase in ECF with load steps is higher for a 30 nm beam compared to the 50 nm beam (Aspect ratio = 12). It is to be noted that, in the absence of geometric nonlinearity, the energy conversion factor is a constant value and remains independent of the applied loads, whereas the consideration of geometric nonlinearity has led to change in ECF value with load increments. 

\begin{figure} \begin{center} 
\subfigure[]{\includegraphics[scale=0.35]{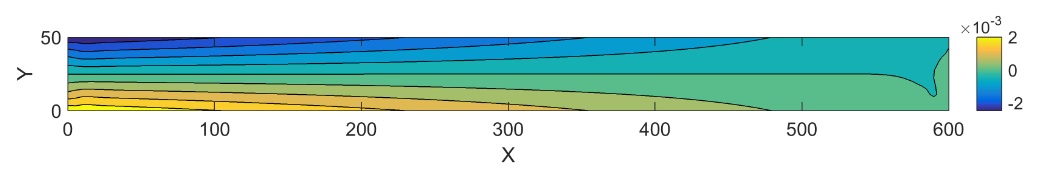}}
\subfigure[]{\includegraphics[scale=0.375]{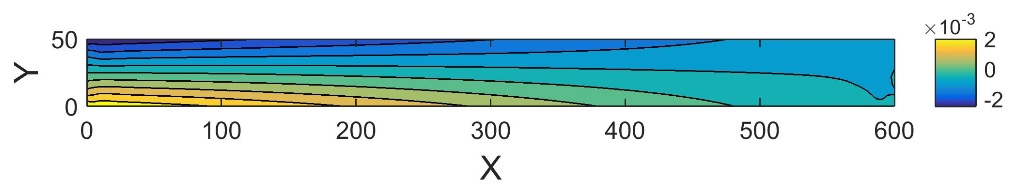}}
\subfigure[]{\includegraphics[scale=0.375]{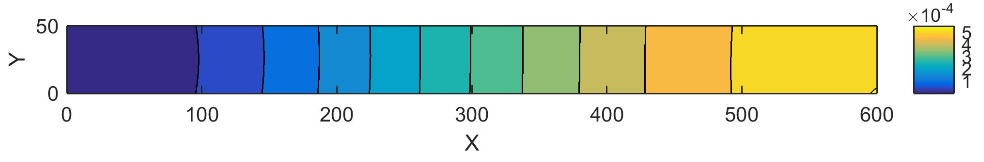}}
\caption{The strain contour ($G_{11}$,$\varepsilon_{11}$,$\eta_{11}$) for flexoelectric beam made of STO of depth 50 nm at load step 1 (a) Green Lagrange strain (b) Linear strain (c) Nonlinear part of strain.}
\label{fig:strain_contour1}
\end{center}
 \end{figure}

\begin{figure} \begin{center} 
\subfigure[]{\includegraphics[scale=0.35]{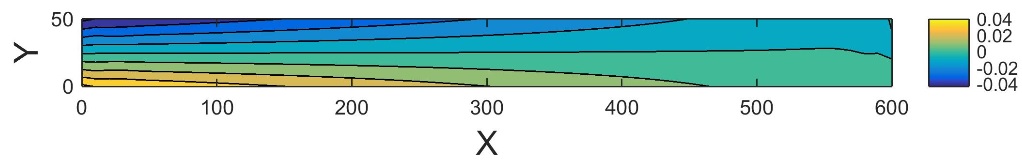}}
\subfigure[]{\includegraphics[scale=0.35]{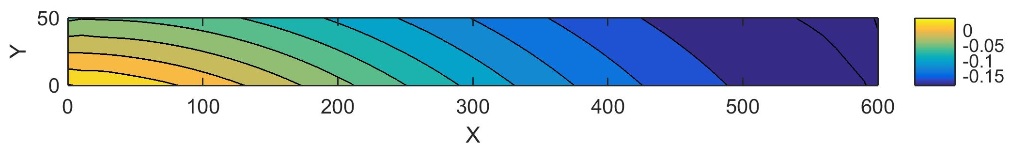}}
\subfigure[]{\includegraphics[scale=0.35]{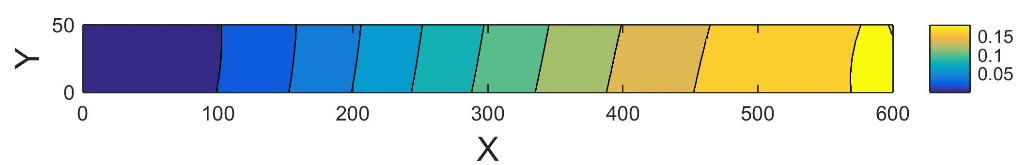}}
\caption{The strain contour ($G_{11}$,$\varepsilon_{11}$,$\eta_{11}$) for flexoelectric beam made of STO of depth 50 nm at load step 10 (a) Green Lagrange strain (b) Linear strain (c) Nonlinear part of strain.}
\label{fig:strain_contour15} \end{center} 
\end{figure}

The contour plot showing the normal strain in x-direction is given in Figures~\ref{fig:strain_contour1} and~\ref{fig:strain_contour15}. Figure~\ref{fig:strain_contour1} shows that at the first load step, the nonlinear strain terms are negligible and the Green-Lagrange strain contour and linear strain contour are quite similar. While the strain contour at the 10$^{th}$ load step given in Figure~\ref{fig:strain_contour15} shows that nonlinear terms play  significant role in determining the Green-Lagrange strain. Consequently, the linear strain contour and Green Lagrange strain contours become dissimilar. The studies performed in this section show that the surface elasticity can reduce the output flexoelectric voltage. For instance, it is observed that in case of a 30 nm thick beam the reduction in the final voltage is 1 mV due to surface elastic effects (Figure~\ref{fig:nonlinear}). The surface elastic effects can increase with increase in the surface area of the beam. 
  two-dimensional model with plane strain assumption, the length of top and bottom face of the beam can be increased by  modifying the rectangular beam into a tapered beam. 
Therefore it is worthwhile to study a tapered beam model with inclined top and bottom faces and compare their response with the rectangular beams studied previously.

\begin{figure} \begin{center} 
\subfigure[]{{\includegraphics[scale=0.5]{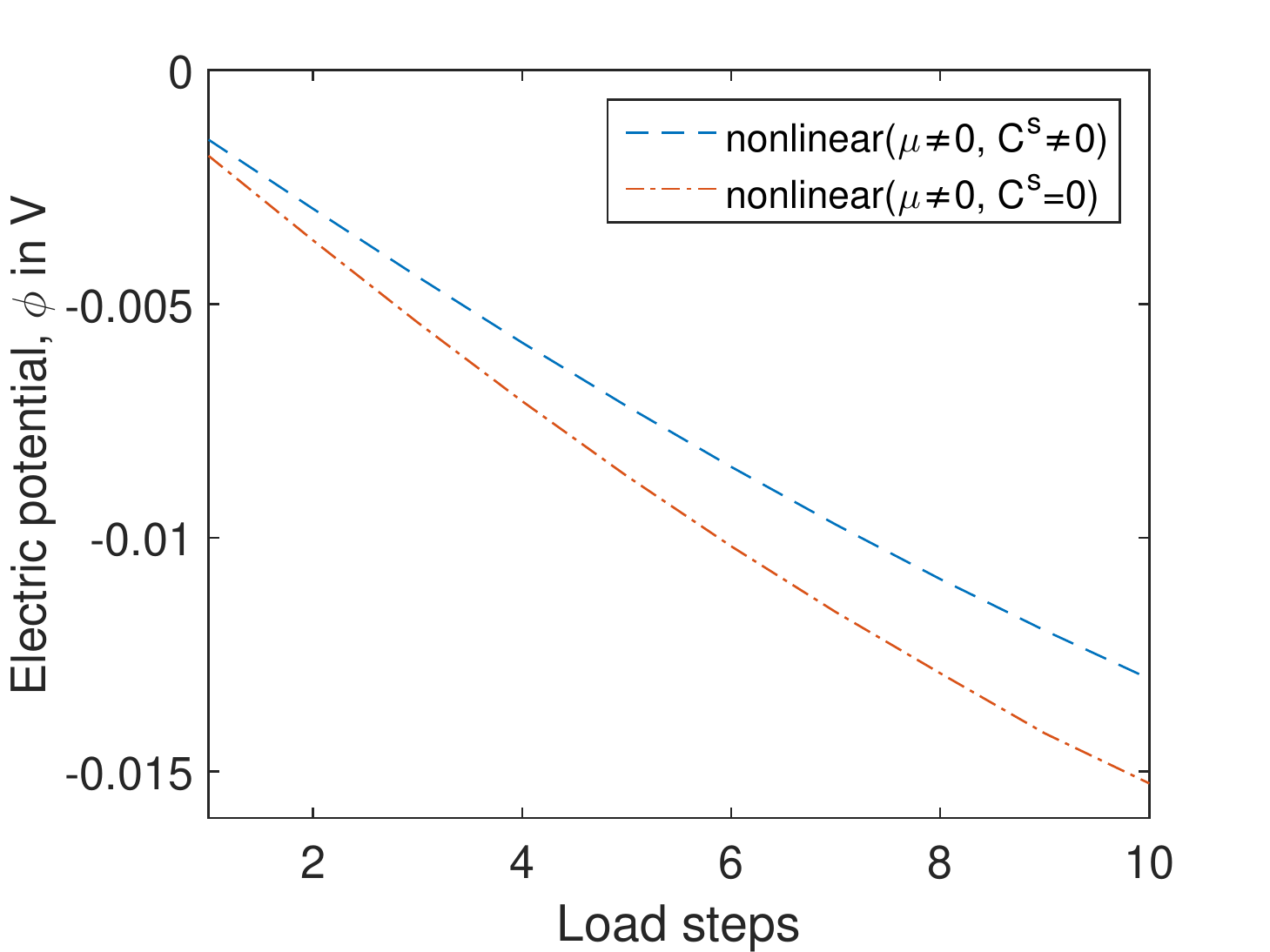}}\label{fig:TB1}}
\subfigure[]{{\includegraphics[scale=0.5]{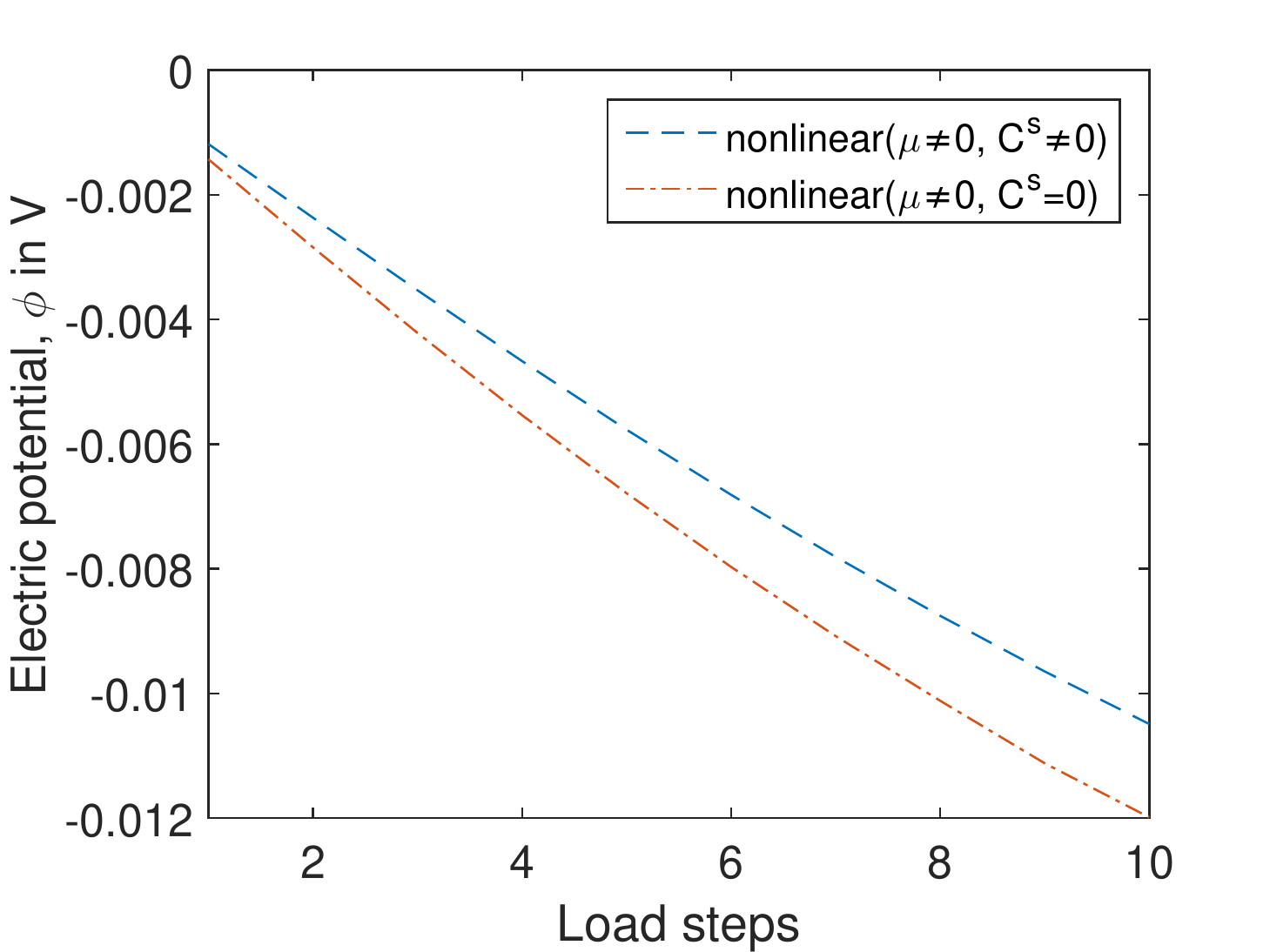}}\label{fig:TB2}}
\caption{The variation in voltage with load increments for flexoelectric tapered beam made of STO of type (a) $TB_1$ and (b) $TB_2$.}
 \end{center} \end{figure}

Considering an average thickness of 30 nm, a tapered cantilever beam, $TB_1$, of length 360 nm and of depth, $d_l$=40 nm at x=0, $d_r$=20 nm at x=360 nm is subjected to a load increment of 1 nN over ten load steps. At the tenth load step, the output flexoelectric voltage in $TB_1$ reduces from 15 mV to 13 mV  due to surface elastic effects as shown in Figure~\ref{fig:TB1}. Secondly, a tapered cantilever beam $TB_2$ of length 360 nm and of depth, $d_l$=45 nm at x=0, $d_r$=15 nm at x=360 nm is considered. The final voltage reduces from 12 mV to 10.4 mV as shown in Figure~\ref{fig:TB2}. In both the tapered beams $TB_1$ and $TB_2$ with average thickness of 30 nm, the voltage reduces by $13.3\%$ due to surface elastic effects. This is higher than the reduction of $8.5\%$ (from 11.8 mV to 10.8 mV) in 30 nm thick rectangular beam. Therefore, as expected the influence of surface effects increases with increase in surface area of the beam and the negative influence on flexoelectric voltage is
  higher for a tapered beam compared to a rectangular beam of same volume.

In summary, the results obtained in this section show that considering nonlinear terms in strain and gradient of strain, is more essential as the influence of flexoelectricity on output voltage gets significant in nanoscale. We conclude that the geometric nonlinearity cannot be ignored if one analyses flexoelectric beams of dimensions of under 100 nanometers when subjected to loads in the range of 10 nNs. It is to be noted that the flexoelectric material, STO used in this example has a lesser flexoelectric constant of only 1.4 V, while flexoelectric constant of a dielectric material can even be upto 10 V based on the theoretical upper limit estimated by Kogan \textit {et al.} \cite{Kogan}.
\\
\\

\begin{figure} \begin{center} 
\includegraphics[scale=0.25]{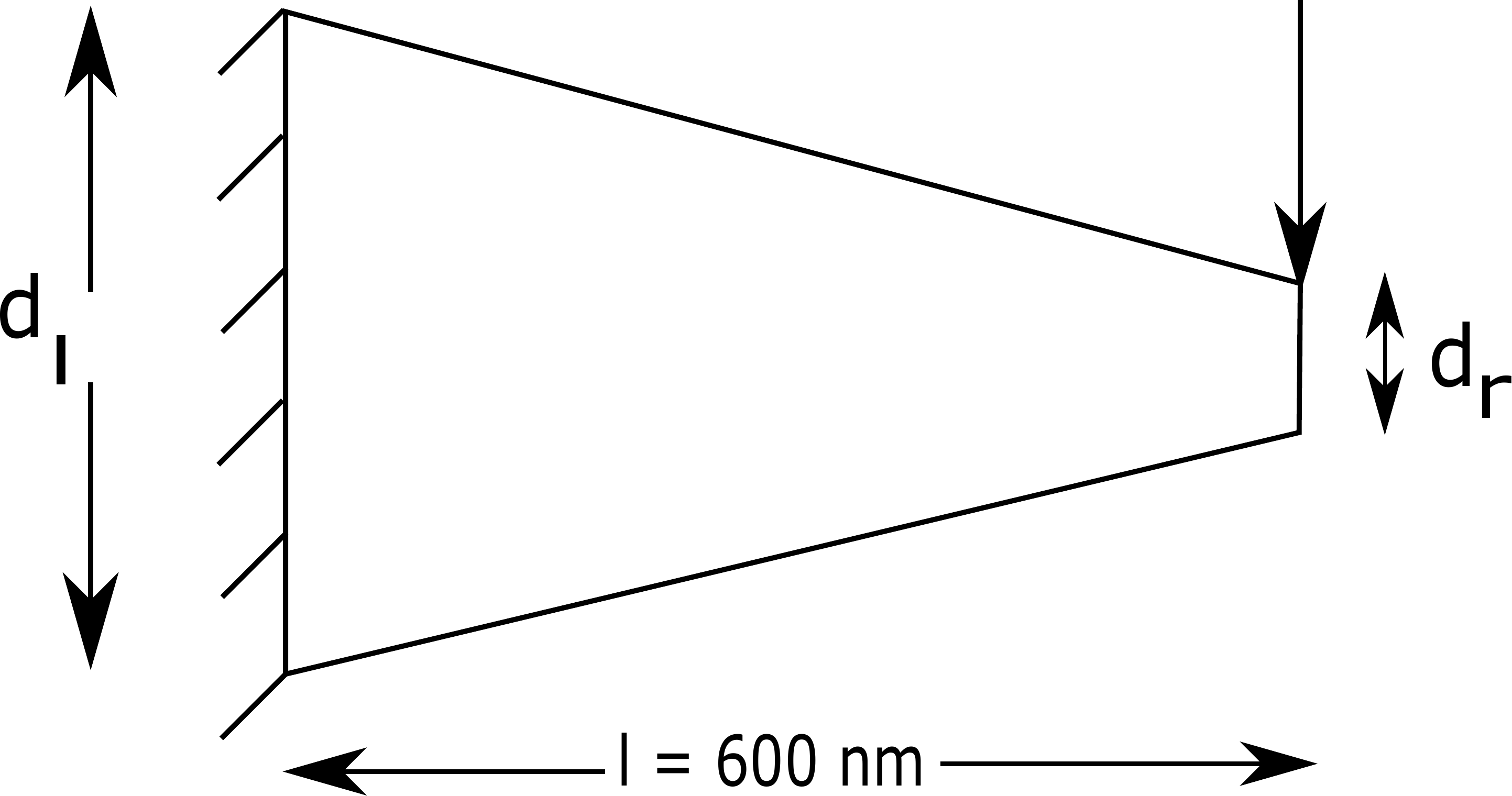}
\caption{Tapered flexoelectric beam model.}
\label{fig:tapered}
 \end{center} \end{figure}
 
\section{Conclusion}
A CSRBF based meshfree formulation is presented in this paper to handle geometric nonlinearity in flexoelectric structures. In addition to flexoelectricity, the surface effects are also considered in the analysis of nano-sized two dimensional structures. Flexoelectric beams made of cubic STO, which is non-piezoelectric and ZnO, which is piezoelectric are analysed. The meshfree analysis shows that for ZnO, the contribution of surface effects to the output voltage of a nanosized cantilever structure (of width 15 nm) is smaller compared to the contribution of flexoelectricity.\\
The analysis of flexoelectric nanostructures undergoing large deformation shows that the difference between nonlinear and linear flexoelectric voltage increases with reduction in beam depth. The surface elastic effects stiffen the beam leading to reduction in output flexoelectric voltage. The influence of surface elasticity is higher in tapered beams compared to rectangular beams. In future, the presented formulation will be extended to study nonlinear flexoelectricity under dynamic excitations and the influence of nonlinearity on response bandwidth of nanosized flexoelectric energy harvesters will be investigated.\\
\appendix
\section{Intermediate Steps in the derivation of nonlinear meshfree formulation for flexoelectricity}
 \label{app1}
The terms $\bm{G}$, $\bm{\delta G}$, $\bm{\Delta \delta G}$ and $\bm{\Delta S}$ in Equation~\ref{eq:LT1} are as follows,
\begin{equation}
\bm{G} = \frac{1}{2}(u_{i,j}+u_{j,i}+u_{k,i}u_{k,j})
\end{equation}
\begin{equation}
\bm{\delta G} = \frac{1}{2}(\delta u_{i,j}+\delta u_{j,i}+\delta u_{k,i}u_{k,j}+u_{k,i}\delta u_{k,j})
\end{equation}

\begin{equation}
\begin{split}
\Delta\bm{\delta G} &= \frac{1}{2}(\delta u_{k,i}\Delta u_{k,j}+\Delta u_{k,i}\delta u_{k,j})\\
                &= \delta u_{k,i}\Delta u_{k,j}\\
                \end{split}
\end{equation}
\begin{equation}
\begin{split}
\bm{S}&=\bm{C}:\bm{G}-\bm{e}\cdot\bm{E}\\
\Delta\bm{S}&=\bm{C}:\Delta\bm{G}-\bm{e}\cdot\Delta\bm{E}\\
\end{split}
\end{equation}
The terms $\bm{\tilde{G}}$, $\bm{{\delta \tilde{G}}}$, $\Delta\bm{{\delta \tilde{G}}}$ and $\bm{\tilde{S}}$ in Equation~\ref{eq:LT2} are as follows,
\begin{equation}
\begin{split}
\bm{\tilde{G}} &= \frac{1}{2}(u_{i,jk}+u_{j,ik}+u_{k,ij}u_{k,j}+u_{k,i}u_{k,ji})\\
\bm{\delta \tilde{G}} &= \frac{1}{2}(\delta u_{i,jk}+\delta u_{j,ik}+\delta u_{k,ij}u_{k,j}+ u_{k,ij}\delta u_{k,j}+\delta u_{k,i}u_{k,ji}+u_{k,i}\delta u_{k,ji})\\
\bm{\Delta \delta \tilde{G}} &= \frac{1}{2}(\delta u_{k,ij}\Delta u_{k,j}+ \Delta u_{k,ij}\delta u_{k,j}+\delta u_{k,i}\Delta u_{k,ji}+\Delta u_{k,i}\delta u_{k,ji})\\
&= \delta u_{k,ij}\Delta u_{k,j}+ \Delta u_{k,ij}\delta u_{k,j}\\
\bm{\tilde{S}}&=-\bm{\mu}\cdot\bm{E}\\
\end{split}
\end{equation}
The terms $\Delta\bm{D}$, $\bm{\delta E}$ and $\Delta\bm{\delta E}$ in Equation~\ref{eq:LT3} are as follows,
\begin{equation}
\begin{split}
\bm{D}&=\bm{e}:\bm{G}+\bm{\mu}\vdots\bm{\tilde{G}}+\bm{\kappa}\cdot\bm{E}\\
\Delta\bm{D}&=\bm{e}:\Delta\bm{G}+\bm{\mu}\vdots\Delta\bm{\tilde{G}}+\bm{\kappa}\cdot\Delta\bm{E}\\
{E_i}&=-\phi_{,i}\\
\delta E_i &= -\delta \phi_{,i}\\
\Delta \delta E_i &= 0\\
\end{split}
\end{equation}
The terms $\bm{G^s}$, $\bm{\delta G^s}$, $\Delta\bm{\delta G^s}$ and $\Delta\bm{S^s}$ in Equation~\ref{eq:LT4} are as follows,
\begin{equation}
\bm{G^s} = \frac{1}{2}P_{ij}(u_{i,j}+u_{j,i}+u_{k,i}u_{k,j})P_{ji}
\end{equation}
\begin{equation}
\bm{\delta G^s} =  \frac{1}{2}P_{ij}(\delta u_{i,j}+\delta u_{j,i}+\delta u_{k,i}u_{k,j}+u_{k,i}\delta u_{k,j})P_{ji}
\end{equation}

\begin{equation}
\begin{split}
\Delta\bm{\delta G^s} &= \frac{1}{2}P_{ji}(\delta u_{k,i}\Delta u_{k,j}+\Delta u_{k,i}\delta u_{k,j})P_{ji}\\
                &= P_{ji}\delta u_{k,i}\Delta u_{k,j}P_{ji}\\
                \end{split}
\end{equation}
\begin{equation}
\begin{split}
\bm{S^s}&=\bm{C^s}:\bm{G^s}\\
\Delta\bm{S^s}&=\bm{C^s}:\Delta\bm{G^s}\\
\end{split}
\end{equation}
\section{Matrices - nonlinear meshfree formulation for flexoelectricity}
 \label{app2}
The expressions defining the matrices $\bm{B}$, $\bm{B_\phi}$, $\bm{H_1}$, $\bm{H_2}$, $\bm{H_u}$, $\bm{H_D}$, $\bm{\hat{R}}$, $\bm{R}$, $\bm{\hat{R}_D}$, $\bm{R_D}$, $\bm{\hat{D}}$, $\bm{{R}_s}$, $\bm{\hat{R}_s}$ and $\bm{P_n}$ are as follows,
\begin{equation}
\bm{B} = \left[\begin{array}{c c}
N_{I,x} & 0 \\ 0 & N_{I,y}\\ N_{I,y} & N_{I,x}
\end{array}\right]+ \bm{A}\,\bm{H_1}
\end{equation}\\
\begin{equation}
\bm{A} = \left[\begin{array}{c c c c}
\frac{\partial u_{Ix}}{\partial x} & 0 & \frac{\partial u_{Iy}}{\partial x} & 0\\
0 & \frac{\partial u_{Ix}}{\partial y} & 0 & \frac{\partial u_{Iy}}{\partial y} \\ \frac{\partial u_{Ix}}{\partial y} & \frac{\partial u_{Ix}}{\partial x} & \frac{\partial u_{Iy}}{\partial y} & \frac{\partial u_{Iy}}{\partial x}
\end{array}\right]
\end{equation}
\begin{equation}
\bm{H_1} = \left[\begin{array}{c c}
N_{I,x} & 0 \\ N_{I,y} & 0 \\ 0 & N_{I,x} \\ 0 & N_{I,y}
\end{array}\right]
\end{equation}
\begin{equation}
\bm{B_\phi}=  \left[\begin{array}{c}
N_{I,x}  \\ N_{I,y}
\end{array}\right]
\end{equation}
\begin{equation}
\bm{H_D} = \bm{H_u}+ \bm{A_D}\,\bm{H_2}
\end{equation}
\begin{equation}
\bm{A_D} = \left[\begin{array}{c c c c c c c c}
\frac{\partial u_{Ix}}{\partial x} & 0 & 0 & 0 & \frac{\partial u_{Iy}}{\partial x} & 0 & 0 & 0 \\
0 & \frac{\partial u_{Ix}}{\partial x} & 0  & 0  & 0 & \frac{\partial u_{Iy}}{\partial x} & 0  & 0  \\  0  & 0 &\frac{\partial u_{Ix}}{\partial y} & 0  & 0  & 0  & \frac{\partial u_{Iy}}{\partial y} & 0 \\
 0  & 0  & 0  & \frac{\partial u_{Ix}}{\partial y}  & 0  & 0  & 0 & \frac{\partial u_{Iy}}{\partial y}\\
 \frac{\partial u_{Ix}}{\partial y} & 0 & \frac{\partial u_{Ix}}{\partial x} & 0 & \frac{\partial u_{Iy}}{\partial y}  0 & \frac{\partial u_{Iy}}{\partial x} & 0 \\
 0 & \frac{\partial u_{Ix}}{\partial y} &  0 & \frac{\partial u_{Ix}}{\partial x} & 0 & \frac{\partial u_{Iy}}{\partial y} & 0 & \frac{\partial u_{Iy}}{\partial x} 
\end{array}\right]
\end{equation}
\begin{equation}
\bm{H_2} =  \left[\begin{array}{c c}
N_{I,xx} & 0\\N_{I,xy} & 0\\N_{I,yx} & 0\\N_{I,yy} & 0\\
0 & N_{I,xx}\\0 & N_{I,xy} \\ 0 & N_{I,yx} \\ 0  & N_{I,yy}
\end{array}\right]
\end{equation}
\begin{equation}
\bm{R} = \left[\begin{array}{c c c c}
S_{11} & S_{12} & 0 & 0 \\ S_{12} & S_{22} & 0 & 0 \\
0 & 0 & S_{11} & S_{12} \\ 0 & 0 & S_{12} & S_{22}
\end{array}\right]
\end{equation}
\begin{equation}
\bm{\hat{R}}=\left[\begin{array}{c c c}
S_{11} & S_{22} & S_{12}
\end{array}\right]^T
\end{equation}
\begin{equation}
\bm{R_D} = \left[\begin{array}{c c c c}
S_{111} & S_{121} & 0 & 0 \\ S_{122} & S_{212} & 0 & 0 \\
S_{121} & S_{122} & 0 & 0 \\ S_{212} &  S_{222} & 0 & 0 \\
0 & 0 & S_{111} & S_{121} \\ 0 & 0 & S_{211} & S_{212} \\
0 & 0 & S_{121} & S_{122} \\ 0 & 0 & S_{212} & S_{222} 
\end{array}\right]
\end{equation}
\begin{equation}
\bm{\hat{R}_D}=\left[\begin{array}{c c c c c c}
S_{111} & S_{211} & S_{122} & S_{222}& S_{121}& S_{212}
\end{array}\right]^T
\end{equation}
\begin{equation}
\bm{\hat{D}} = \left[\begin{array}{c}
D_1 \\ D_2 \end{array}\right]
\end{equation}
\begin{equation}
\bm{\hat{R}_s}=\left[\begin{array}{c c c}
\tau_{11}^s+S^s_{11} & \tau_{22}^s+S^s_{22} & \tau_{12}^s+S^s_{12}
\end{array}\right]^T
\end{equation}
\begin{equation}
\bm{R_s} = \left[\begin{array}{c c c c}
\tau_{11}^s+S^s_{11} & \tau_{12}^s+S^s_{12} & 0 & 0 \\ \tau_{12}^s+S^s_{12} & \tau_{22}^s+S^s_{22} & 0 & 0 \\
0 & 0 & \tau_{11}^s+S^s_{11} & \tau_{12}^s+S^s_{12} \\ 0 & 0 & \tau_{12}^s+S^s_{12} & \tau_{22}^s+S^s_{22}
\end{array}\right] 
\end{equation}
\begin{equation}
\bm{P_n} = \left[ \begin{array}{cccc} P_{11} &P_{12} &0 & 0 \\ P_{21} &P_{22} &0 & 0 \\0&0&P_{11} &P_{12} \\ 0&0&P_{21} &P_{22} 
\end{array}
\right]
\end{equation}
\section*{Acknowledgments}
X. Zhuang acknowledges the support from State Key Laboratory of Structural Analysis for Industrial Equipment (GZ1607) and National Science Foundation of China (11772234).\\ 
\\
\section*{References}
\bibliographystyle{model1-num-names}
\bibliography{Reference}
\end{document}